\documentclass[a4paper,11pt]{article}
\pdfoutput=1
\usepackage{jcappub}
\usepackage{amsthm,graphicx,multirow}
\usepackage{epsfig}
\usepackage{latexsym, amssymb} 
\usepackage{amsmath}
\def\beq{\begin{equation}}
\def\eeq{\end{equation}}
\def\br{\begin{eqnarray}}
\def\er{\end{eqnarray}}
\def\benu{\begin{enumerate}}
\def\efnu{\end{enumerate}}
\def\nn{\nonumber}

\def\l{\left}
\def\r{\right}

\def\d{{\rm d}}

\def\vka{{\bf k}_{1}}
\def\vkb{{\bf k}_{2}}
\def\vkc{{\bf k}_{3}}

\def\cB{{\cal B}}
\def\fnl{f_{_{\rm NL}}}

\begin{document}
\begin{center}
\title{Primordial features and Planck polarization}
\end{center}
\author[a]{Dhiraj Kumar Hazra,} 
\author[b,c]{Arman Shafieloo,}
\author[a,d,e]{George F. Smoot,}
\author[f,g]{Alexei A. Starobinsky}

\affiliation[a]{AstroParticule et Cosmologie (APC)/Paris Centre for Cosmological Physics, Universit\'e
Paris Diderot, CNRS, CEA, Observatoire de Paris, Sorbonne Paris Cit\'e University, 10, rue Alice Domon et Leonie Duquet, 75205 Paris Cedex 13, France}
\affiliation[b]{Korea Astronomy and Space Science Institute, Daejeon 34055, Korea}
\affiliation[c]{University of Science and Technology, Daejeon 34113, Korea}
\affiliation[d]{Institute for Advanced Study, Hong Kong University of Science and Technology, Clear Water Bay, Kowloon, Hong Kong}
\affiliation[e]{Physics Department and Lawrence Berkeley National Laboratory, University of California, Berkeley, CA 94720, USA}
\affiliation[f]{Landau Institute for Theoretical Physics RAS, Moscow, 119334, Russian Federation}
\affiliation[g]{Kazan Federal University, Kazan 420008, Republic of Tatarstan, Russian Federation}
\emailAdd{dhiraj.kumar.hazra@apc.univ-paris7.fr, shafieloo@kasi.re.kr, gfsmoot@lbl.gov, alstar@landau.ac.ru} 

\abstract 
{With the Planck 2015 Cosmic Microwave Background (CMB) temperature and
polarization data, we search for possible features in
the primordial power spectrum (PPS). We revisit the Wiggly Whipped
Inflation (WWI) framework and demonstrate how generation of some
particular primordial features can improve the fit to Planck data. WWI
potential allows the scalar field to transit from a steeper potential
to a nearly flat potential through a discontinuity either in potential
or in its derivatives. WWI offers the inflaton potential parametrizations that 
generate a wide variety of features in the primordial power spectra incorporating 
most of the localized and non-local inflationary features that are obtained upon 
reconstruction from temperature and polarization angular power spectrum. At 
the same time, in a single framework it allows us to have a background parameter 
estimation with a nearly free-form primordial spectrum. Using Planck 2015 data, we constrain the
primordial features in the context of Wiggly Whipped Inflation and
present the features that are supported both by temperature and
polarization. WWI model provides more than $13$ improvement in
$\chi^2$ fit to the data with respect to the best fit power law model
considering combined temperature and polarization data from Planck and
B-mode polarization data from BICEP and Planck dust map. We use
2-4 extra parameters in the WWI model compared to the featureless
strict slow roll inflaton potential. We find that the differences between the temperature and
polarization data in constraining background cosmological parameters
such as baryon density, cold dark matter density are reduced to a good
extent if we use primordial power spectra from WWI. We also discuss
the extent of bispectra obtained from the best potentials in arbitrary
triangular configurations using the BI-spectra and Non-Gaussianity
Operator (BINGO).}

\maketitle
\section{Introduction}

While viable inflationary models of the early Universe have succeeded in the quantitative description of the main, smooth part of the primordial
power spectrum (PPS) of scalar (adiabatic density) perturbations $P_{\rm S}(k)$, and have even predicted its observed slope $n_{\rm S}(k)$ 
for some simplest variants of them, the complete shape of the primordial power spectra has not been established beyond doubt. 
It is natural to expect small corrections to this smooth behavior which reflect new and subtle physical effects occurring during inflation. Indeed, some 
relatively small features ($\lesssim 10\%$) have already been noticed in the low-$\ell$ ($\ell\lesssim 40$) multipole region of CMB fluctuations since 
WMAP, which may result from tiny localized features in the PPS.

With the release of the new Planck temperature and polarization data~\cite{Planck:2015Like,Planck:2015Param,PLA}, we are now in a unique position to examine the existence 
of primordial features in the scalar perturbations to a great precision over a wide range of cosmological scales. We had indication 
of primordial features in all releases of WMAP temperature data~\cite{Peiris:2003ff,Komatsu:2008hk,Komatsu:2010fb,Hinshaw:2012aka} and Planck 2013-2015 temperature data~\cite{Planck:2013Inf,Planck:2015Inf}
and we have knowledge about the location and the types of the features from reconstructions of the primordial power spectrum. Inflaton potentials, 
addressing the primordial features were proposed along the lines, that provided notable, if not significant, improvement in fit compared to the standard power law 
primordial spectrum from slow roll inflation. A large scale scalar suppression, a dip near $2\times10^{-3} {\rm Mpc^{-1}}$, oscillations around 
$0.02~{\rm Mpc^{-1}}$ and around $0.06~{\rm Mpc^{-1}}$ wavenumbers, are few notable features from Planck 2013 temperature anisotropy data~\cite{Hazra:2014PPSPlanck}.
Interestingly, the necessity of the large scale suppression of scalar power became significant when BICEP2 B-mode signal were considered primordial~\cite{BICEP2:Detection,BICEP2:datasets}.
The effect of tensor perturbation at the large scale temperature anisotropies aggravated (at more than 3$\sigma$ level~\cite{Hazra:2014RuleOut}) the already existing issue of lack of 
low-multipole power of temperature anisotropies (1$\sigma$ support of power suppression from Planck 2013 TT data)~\cite{Hazra:2013BroadRecon}. We proposed Whipped Inflation (WI)~\cite{Hazra:2014WI} and Wiggly 
Whipped Inflation (WWI)~\cite{Hazra:2014WWI} that could address all the above mentioned issues with the primordial power spectra and provided significant improvement in fit compared to 
canonical slow roll models of inflation. Today, with the BICEP2 signal being consistent to dust polarization, one can anticipate a dramatic decrease in significance
of the large scale suppression. Also the large magnitude of quadratic potential, used in WI and 
WWI makes it unsuitable for recent BICEP2/KECK-Planck joint constraints on the primordial B-modes (tensor-to-scalar ratio $r<0.07$ at 95\% confidence level~\cite{BICEP2Planck:2015joint,Array:2015xqh}).
However, note that the significance of the Wiggles in the WWI are not altered by change in the BICEP2 results since they were designed to address features 
in Planck temperature data. 

Since we now have tighter constraints in background and primordial cosmology from the new temperature and polarization data, it is important to revisit 
the status of features in the light of new data. It is also appropriate to modify the WWI potential that can satisfy present bounds on B-mode and dust polarization.
If temperature and polarization votes for similar features it definitely shall increase its likelihood, hence equipped with Planck polarization data covering the largest 
cosmological scales it is definitely worth having a joint constraints on features. Change in constraints on the background cosmological parameters in the presence
of features is another aspect that should be investigated. In this paper, we explore the above mentioned. It is clearly not possible to examine every primordial feature within 
a single framework of potential. However, we show that our WWI potential is capable of producing a large variety of features that are discussed in the literature, which makes it extremely
efficient to hunt different types of features required by different datasets in the same framework. 

The paper is organized as follows: In section~\ref{sec:scenario} we review the WWI potential in its old and modified form. We demonstrate to what extent WWI potential 
can offer a variety of features in the PPS. In section~\ref{sec:num} we discuss the methodology, {\it i.e.} a brief outline of our numerical analysis. We provide our best fit results and
constraints on the model in section~\ref{sec:results}. We also present the bispectra from the best fits to the data in arbitrary triangular configurations of wavenumbers. We conclude 
in section~\ref{sec:conclusions}. 

\section{Wiggly Whipped Inflationary Scenario}~\label{sec:scenario}

The basic construction of the Wiggly Whipped Inflation potential followed the form :
\begin{equation}
V({\phi})=V_{S}+\gamma V_{R},~\label{eq:equation-WWI-basic}
\end{equation}
where, $V_{R}$ represents the part of the potential that introduces a moderate fast roll and $V_{S}$ is chosen to be nearly flat, allowing a strict slow roll. 
Here, $\gamma$ determines how fast the inflaton starts rolling initially and in a way it determines the extent of deviation from slow-roll and thereby
deviations from featureless PPS. Similar types of phase transition potential have been discussed widely in the literature~\cite{S92,Linde:1998OpenInf,Linde:1999OpenInf2,CPKL03,JSS08,JSSS09,Bousso:2013uia}. 
$V_{R}$ contains a Theta step function ensuring the termination of moderate fast roll after certain field value. Thereafter the scalar field emerges to the strict slow roll regime.  
In this framework, the potential and/or its derivatives can contain discontinuities.

The WWI model was introduced in~\cite{Hazra:2014WWI} in order to address two major issues. Firstly with the BICEP2 B-mode 
signal, assuming to be primordial, we ruled out the power law form of scalar power spectrum with more than 3$\sigma$ confidence~\cite{Hazra:2014RuleOut}. We needed 
a strong suppression of scalar power at large scales and at the same time large tensors with low non-Gaussianities. Whipped inflation~\cite{Hazra:2014WI} that 
simply uses a smooth transition from a moderate fast roll potential ($V_{R}$) to a strict slow roll inflation ($V_{S}$, assumed to be a quadratic potential) meets 
all the above criteria. On top of that in order to address the primordial features indicated by Planck temperature anisotropies, we introduced a discontinuity in the potential (or in its 
derivatives, keeping a continuous potential) at the transition which provided significant improvement in likelihood to the Planck and assumed primordial B-mode from BICEP2 compared to the 
power law PPS. The scalar PPS, containing
a Whipped shaped tail at large scales and Wiggles (oscillatory behavior), the Inflaton potential was referred as Wiggly Whipped Inflation potential. The BICEP-2 B-mode signal being consistent to dust polarization, 
do not favor a large field quadratic model since it produce large primordial B-modes. At the same time the significance of the requirement of a large scale scalar suppression 
reduces. However, since we now have polarization data from Planck, it is important to test the WWI model features with the new datasets. In this paper we use the 
WWI potential with a modification to the slow roll part of the potential. 

\subsection{The potentials}

We use the WWI model in a modified form. Since BICEP2 B-mode signal is consistent with dust polarization power spectra
from Planck, a model with high tensor-to-scalar ratio ($r\sim0.2$) is not supported by the data anymore. Hence we use a
lower potential in the slow roll part. We use two different classes of WWI potential here, in the first case we have discontinuity
in the potential while in the second case the potential is continuous with a discontinuity in its slope. Both of them belong to the WWI form, but 
to distinguish them conveniently, we refer the first potential as WWI and the second as WWI$'$.
Note that since the tensor perturbations depends on the scale of the potential, in order to get a lower 
tensors we need to scale down the potential. Here, our basic structure of the potential remains same and provided below in Eq.~\ref{eq:equation-WWI} 

\begin{equation}
V({\phi})=V_{i} \l(1-\l(\frac{\phi}{\mu}\r)^{p}\r)+\Theta(\phi_{\rm T}-\phi)\gamma V_{i}\l((\phi_{\rm T}-\phi)^{q}+\phi_{01}^q\r)~\label{eq:equation-WWI}
\end{equation}
The slow roll potential $V_{i} \l(1-\l(\frac{\phi}{\mu}\r)^{p}\r)$ depends on 2 parameters, namely the potential at $\phi=0$ ($V_{i}$) and $\mu$. 
The spectral index ($n_{\rm s}$) and the tensor-to-scalar ratio ($r$) of the PPS generated at the end of inflation depends on $\mu$ and the power $p$. 
We chose the value $p=4$ and $\mu=15~{\rm M_{PL}}$ such that $n_{\rm s}\sim0.96$ and $r\sim{\cal O}(10^{-2})$ (as in~\cite{Efstathiou:2006ak,Hazra:2010Step}). 
The fast roll part of the potential remains identical to the original WWI potential. $\phi_{T}$ denotes the field value where the transition from the moderate 
fast-roll to complete slow-roll occurs. $\phi_{01}$ is the extent of discontinuity and $\gamma$ is the slope that provides the deviation of slow-roll at the onset 
of inflation. $\phi_{01}=0$ reduces the potential to Whipped Inflation form where the potential and its derivatives are continuous upto $(q-1)$'th derivatives.
The Heaviside Theta function $\Theta(\phi_{\rm T}-\phi)$ appearing in the potential can be modeled by a Tanh step ($1+{\tanh}[{(\phi-\phi_{\rm T})}/{\Delta}]$)
or Error function step ($1+{\rm erf}[{(\phi-\phi_{\rm T})}/{\Delta}]$) and the width of the step ($\Delta$) can be used as a free parameter. The scalar field
starts from the bottom of a quadratic potential (for $q=2$) and transits to the strict slow roll potential at $\phi=\phi_{\rm T}$. The initial deviation from slow-roll
introduces a whip shaped suppression of power at large and intermediate scales in the PPS. $\phi_{01}>0$ creates a temporary sharp departure from slow-roll and 
generates wiggly features locally or extending a large range in cosmological scales depending on the sharpness of transition, $\Delta$. Hereafter, throughout the paper, 
with WWI, we shall refer to the potential in Eq.~\ref{eq:equation-WWI}. 

Using similar formalism, we investigate another potential that is continuous, but has discontinuities in its derivatives. The potential can be expressed as, 
\begin{equation}
V({\phi})=\Theta(\phi_{\rm T}-\phi) V_{i} \l(1-\exp\l[-\alpha\kappa\phi\r]\r)+\Theta(\phi-\phi_{\rm T}) V_{ii}\l(1-\exp\l[-\alpha\kappa(\phi-\phi_{01})\r]\r)~\label{eq:equation-WWI'}
\end{equation}
Here $V_{i} \l(1-\exp\l[-\alpha\kappa\phi\r]\r)$ is the slow roll part of the potential which is present at $\phi\le\phi_{\rm T}$ and for higher field values, 
$V_{ii}\l(1-\exp\l[-\alpha\kappa(\phi-\phi_{01})\r]\r)$ represents the moderate fast roll part. 
Note that we have used the $\alpha$-attractor potential~\cite{alpha_attractor} to construct a model providing deviations from slow roll. $V_{ii}$ is related to $V_{i}$ by our demand of continuity in the potential. 
$\kappa^2=8\pi G$ (which we equate to 1 in our convention). $\alpha$ denotes the slope of the potential and we have fixed it to be $\sqrt{2/3}$ that corresponds to the $R+R^2$ inflationary model~\cite{Starobinsky:1980te} 
in the Einstein frame. It produces $r\sim 4\times10^{-3}$ for the slow roll part of the spectra. Hence $\phi_{\rm T}$ and $\phi_{01}$ are the only two extra parameters in our potential compared to the strict slow 
roll part of the potential. Note that this potential can also have a discontinuity if $V_{ii}$ is treated as a free parameter, but since we have already incorporated potential discontinuities in Eq.~\ref{eq:equation-WWI}, in 
Eq.~\ref{eq:equation-WWI'} we fix $V_{ii}$ through the continuity. We denote this potential with WWI$'$ to indicate that the potential has a discontinuity only in its derivatives. 
The primordial feature generated from WWI$'$ is very similar to the original Starobinsky-1992 model~\cite{S92} but here, the scalar PPS has the asymptotic value of $n_{\rm S}\sim0.96$. 

\subsection{Classes of features in WWI}
CMB angular power spectrum data since WMAP have indicated hints towards possible deviations from the standard power law PPS. Reconstructions of the PPS
directly from the angular power spectrum data~\cite{Hazra:2014PPSPlanck,reconstruction-all} have been very useful in order to highlight different 
locations and types of these deviations (features) in the PPS. Using Planck 2013 data, it has been demonstrated~\cite{Hazra:2014PPSPlanck} that the 
temperature anisotropy power spectrum indicates a suppression at large scale power, few localized oscillations around $\ell\sim22$, $\ell\sim250-300$ and $\ell\sim750-850$. 
Non-standard inflationary models have been proposed in order to generate different kinds of features in the PPS. Below we enlist a non-exhaustive discussion of primordial features. 

\begin{figure*}[!htb]
\begin{center} 
\resizebox{210pt}{160pt}{\includegraphics{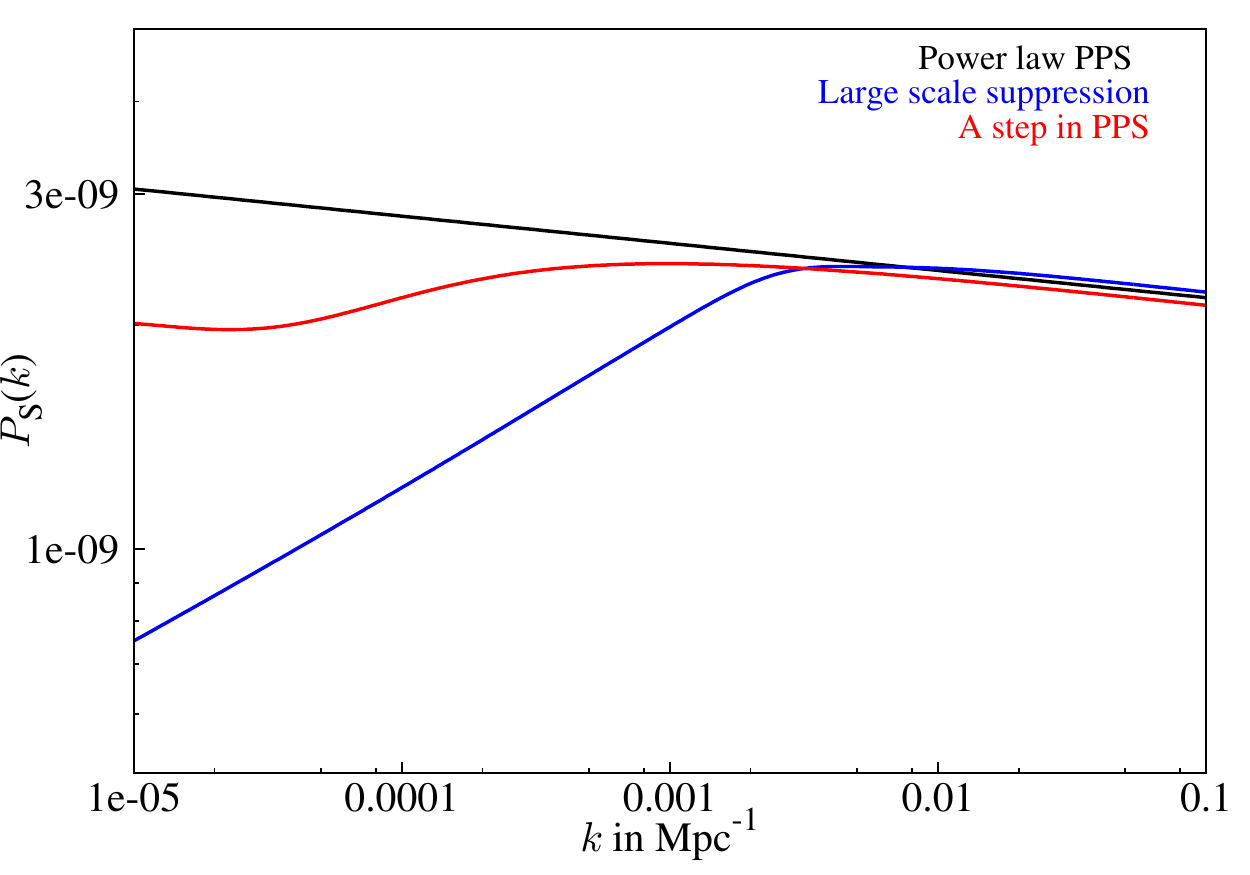}} 
\resizebox{210pt}{160pt}{\includegraphics{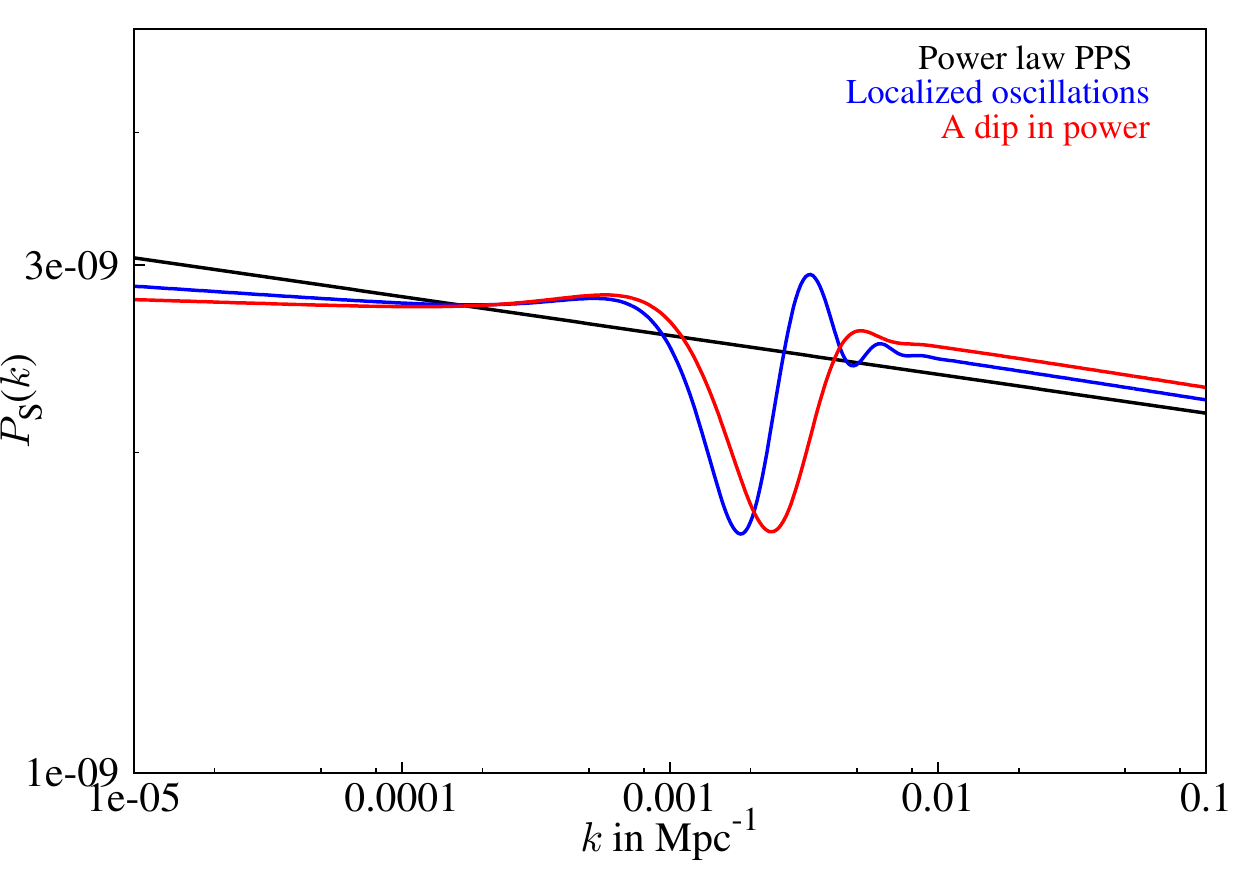}} 

\resizebox{210pt}{160pt}{\includegraphics{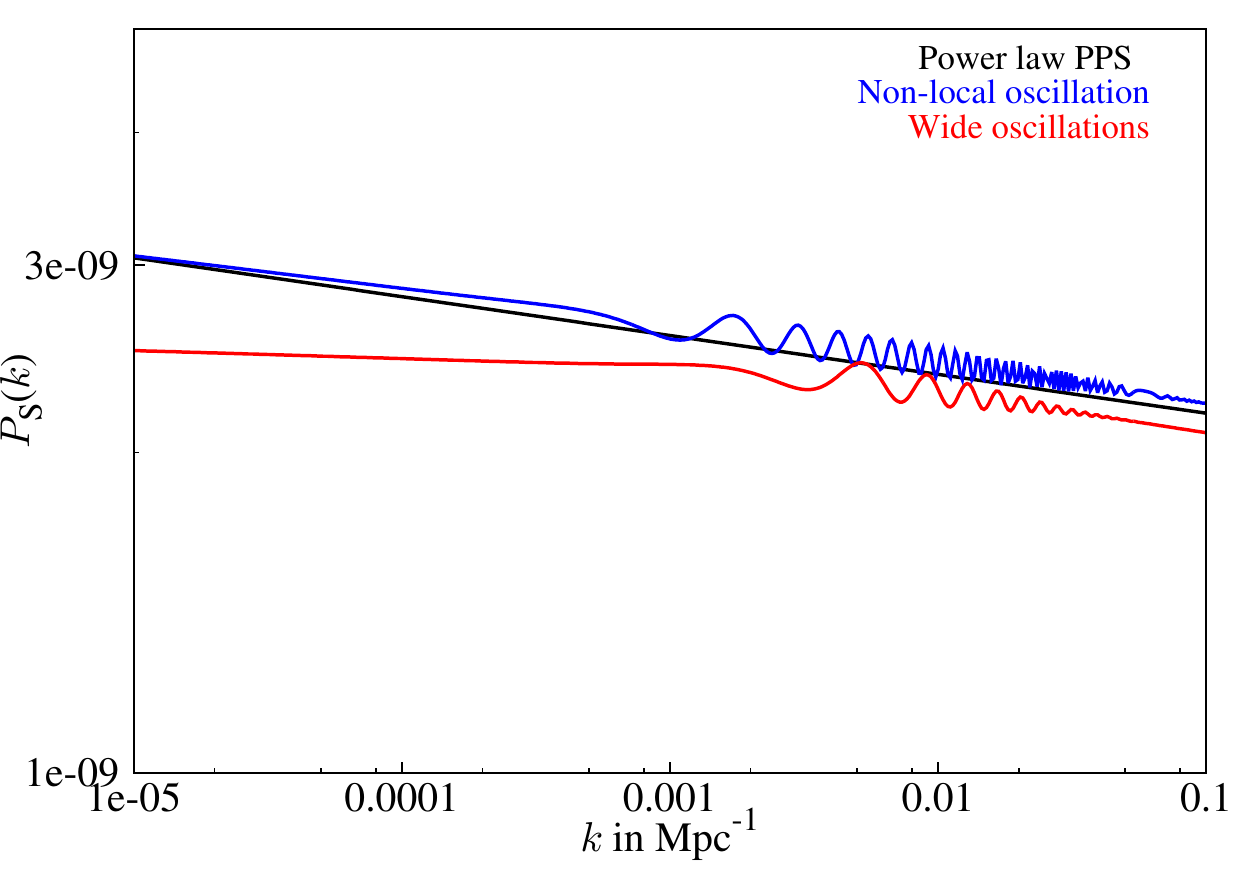}} 
\resizebox{210pt}{160pt}{\includegraphics{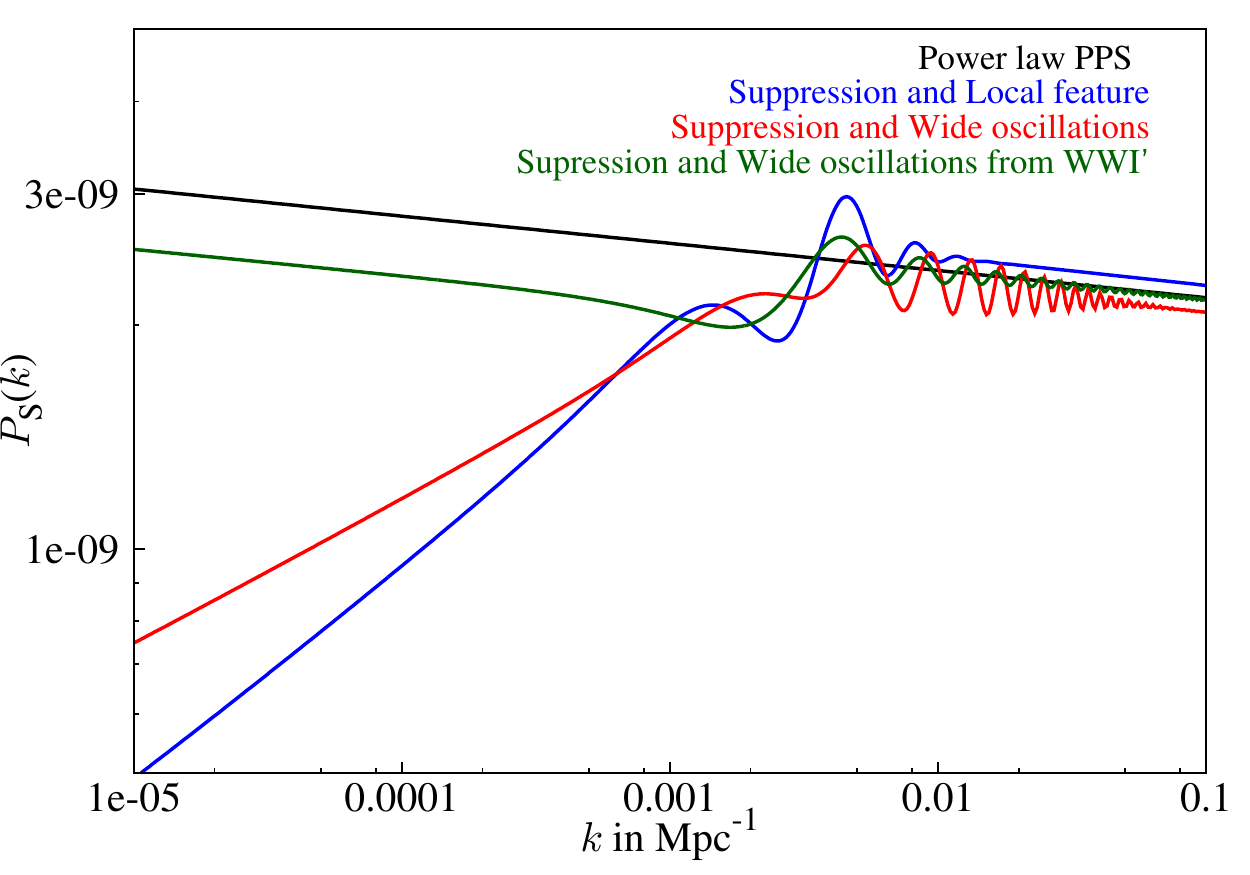}} 

\end{center}
\caption{\footnotesize\label{fig:feature-range} [Top left] : PPS with suppression at large scale scalar power. We show that WWI (Eq.~\ref{eq:equation-WWI}) can generate a exponential cutoff (blue) type 
PPS and also a PPS with a form of a step (red). [Top right] Here WWI produces localized oscillations. Note that a dip or a period of oscillation around $0.002~{\rm Mpc}^{-1}$ 
improves the fit to the data near multipole $\ell=22$. [Bottom left] Wide features and non-local oscillations. The angular power spectra are affected for a wide range of multipoles.
[Bottom right] We provide a combination of different features. The large range of features that are addressed by WWI models, makes WWI a suitable candidate to search for generic 
features in the primordial power spectrum. A typical step like suppression, accompanied by wiggles in the PPS is generated by the WWI$'$ (Eq.~\ref{eq:equation-WWI'}) and is plotted here as well.}
\end{figure*}

1. Large scale power suppression : The primordial power spectrum with a large scale suppression can be modeled by a broken power law with different spectral index at different scales, or 
a Tanh step (both these types have been discussed in~\cite{Hazra:2013BroadRecon}) or an exponential cutoff~\cite{Shafieloo:2004}. These types of features are generated 
when the scalar field changes its kinetic energy in the first few {\it e-folds}. Starting from Starobinsky-1992 model~\cite{S92}, 
different models~\cite{Bousso:2013uia,Bousso:2014jca,Hazra:2014WI} have been 
proposed that offers such scenarios. A brief halt in inflation caused by an inflection point~\cite{Jain:2009PI} in the potential can produce a sharp cut-off at the large scale 
primordial power. Open inflation~\cite{Linde:1998OpenInf,Linde:1999OpenInf2}, radiation dominated epoch prior to inflation~\cite{Powell:2006RADDOM} also known 
to provide such class of spectra. In Fig.~\ref{fig:feature-range}
we plot such type of spectra (in the top left) that are obtained from WWI (Eq.~\ref{eq:equation-WWI}). Note that in these cases, large value of the parameter $\gamma$
allows an initial faster roll and hence provide a cutoff at the large-intermediate scales.

2. Localized oscillation : Near multipoles $\ell\sim22$ and $\ell\sim40$ a dip and a bump {\it w.r.t.} the angular power spectra from power law were noticed since 
WMAP data. This feature patterns in a localized scale and requires wiggles in the PPS only at certain wavenumbers, keeping the other part of the PPS nearly scale
invariant. Presence of {\it a step in the inflaton potential}~\cite{Hazra:2010Step,step-models} provides a momentary departure from the slow roll and generate localized 
oscillations in the PPS. It has been shown that around $k\sim0.002~{\rm Mpc^{-1}}$ and $k\sim0.004~{\rm Mpc^{-1}}$, a period of oscillation can improve the match to the 
data compared to the power law model. However, because of uncertainties owing to the cosmic variance, the existence of such large scale feature 
have never been (possibly never will be) established beyond doubt. Top right of Fig.~\ref{fig:feature-range} shows similar features from WWI. Here, $\gamma$ is small such that 
the fast roll part is not significant and we get nearly same tilt both at large and small scales. However, due to $\phi_{01}$, the field goes through a momentary
departure from slow roll and generates localized oscillations.

3. Non-local features : Features that extend over a wide range of cosmological scales are termed as non-local features. A slow roll potential modulated with sinusoidal 
oscillations or presence of discontinuities in the potential and its derivatives give rise to such non local features. Effects of oscillations in the inflation potential~\cite{oip} 
have been studied before with WMAP and Planck datasets. In~\cite{Hazra:2014WWI} we discussed the features generated by the WWI model and demonstrated 
that non-local features in the WWI can provide similar improvement in fit as local feature models. Fig.~\ref{fig:feature-range}, bottom left plot demonstrates such features
from WWI. Here, $\gamma$ and $\phi_{01}$ are small and hence we do not observe any cutoff and the amplitude of the features are small as well. However, in these models $\Delta<<1$
and hence the field experience a sharp transition and a wide range of modes that leaves Hubble scale after the epoch of transition imprints oscillations in the PPS. 

Wiggly Whipped Inflation, interestingly, is capable of generating all the above features and hence, we present it as the model offering a wide variety of 
scenarios, within the framework of canonical scalar field Lagrangian. Fig.~\ref{fig:feature-range} bottom right plot represent two such PPS that offers the aforementioned classes 
of features in combination, all arising from the same potential. Note that in the same plot we present the feature obtained from WWI$'$. The discontinuity in the derivative
of the potential leads to a power spectrum that offers a step shaped suppression at larger scales, followed by wiggles at the smaller scales. In this case, the
amplitude of oscillations decrease as we probe smaller scales.

\section{Essential numerical details}~\label{sec:num}

We use the publicly available code BI-spectra and Non-Gaussianity Operator, {\tt BINGO}~\cite{Hazra:2012BINGO,Sreenath:2014BINGO2} to generate 
the power spectra and the bispectra from WWI and WWI$'$. We solve the background equation using a initial value of the field $\phi$ to ensure enough ($\sim70$ {\it e-folds}) inflation. 
$\d\phi/\d t$ is fixed assuming initial slow-roll condition ($3 H \d\phi/\d t=-\d V(\phi)/\d \phi$). Whenever necessary, we model the Theta function discontinuities 
in the potential and delta function in its derivatives with a Tanh step and its derivatives respectively. Note that other representations of Theta function and 
delta function can also be used in this context. Following standard methodology we fix the initial scale factor by assuming the $k=0.05~{\rm Mpc}^{-1}$ mode leaves 
the Hubble radius 50 {\it e-folds} before the end of inflation.

We use publicly available {\tt CAMB}~\cite{cambsite,Lewis:1999bs} and {\tt COSMOMC}~\cite{cosmomcsite,Lewis:2002ah} in order to calculate the angular power spectra from
our models and compare them with the data. Note that, we modify {\tt CAMB} in order to use {\tt BINGO} to calculate the primordial scalar and tensor power spectra. Along with 
the baryon density ($\Omega_{\rm b} h^2$), cold dark matter density ($\Omega_{\rm CDM} h^2$), the ratio of the sound horizon to the angular diameter distance at decoupling ($\theta$) 
and the optical depth ($\tau$), we allow the WWI parameters, namely
$V_{i}$, $\gamma$, $\phi_{01}$ and $\phi_{\rm T}$ to vary. We treat the WWI parameters as semi slow-parameters (like the amplitude $A_{\rm S}$, spectral index $\rm n_{\rm S}$, 
tensor-to-scalar ratio ($r$) in case of power law PPS). Note that we also allow the width ($\Delta$) of the Theta function to vary along with other potential parameters. 
For WWI$'$ we treat $V_i$ and $\phi_{\rm T}$ and $\phi_{01}$ as potential variables.  
In order to obtain the best fits we use Powell's BOBYQA (Bound Optimization BY Quadratic Approximation) method of iterative minimization~\cite{powell}. Since WWI offers
a wide variety of features, it is extremely difficult to converge to a global minima starting from a particular region of potential parameter space. Most of the times 
the method settles to a local minima which represent particular features that provide a better fit to a subset of the complete Planck datasets. We approach this problem 
in 2 steps. First, we use temperature, polarization datasets separately and in combination so that we can identify the primordial features supported by the individual and 
complete Planck and BICEP2/KECK datasets. Secondly, in each of the above cases we perform MCMC analyses and locate distinguishable features that provide better fit to the 
corresponding datasets compared to power law model. We use the WWI potential parameters and corresponding background cosmological parameters corresponding to the located 
features and use them as starting points of BOBYQA minimization. Using this rigorous search, we are able to obtain the local minima and possibly the global minima for the individual and
complete datasets.

We use CMB temperature and polarization data from Planck-2015 public release datasets and likelihoods. In order to understand the necessity of features in the PPS indicated from temperature
and polarization data, we use the temperature and polarization likelihood from Planck separately and in combination. For high-$\ell$ temperature and polarization data,
we use the Plik likelihood that covers the multipoles $\ell=30-2508$ for TT and $30-1996$ for TE and EE. 

For the low-$\ell$ part of the spectra ($\ell=2-29$) we use 2 different likelihoods in different cases. When we use only TT likelihood at high-$\ell$, we use 
commander based likelihood for low-$\ell$ TT. In this paper we denote this likelihood as lowT. We use lowTEB likelihood at low-$\ell$ whenever we use EE data at high-$\ell$
or temperature and polarization likelihood in combination. The low-$\ell$ polarization uses the 70GHz LFI full mission (except second and fourth surveys) data. We use 
TTTEEE likelihood and use TT + TE + EE likelihood at high-$\ell$ to track down the improvement in likelihood (compared to the power law model of the PPS) in 
complete and individual datasets. The high-$\ell$ likelihood uses 100, 143 and 217 GHz half mission maps. Throughout our analyses, we have varied all the required 
nuisance and calibration parameters for the corresponding likelihoods. We also use priors on nuisance and calibration parameters as have been discussed in~\cite{Planck:2015Like,PLC}. 

Since the WWI and WWI$'$ potentials generate fine oscillations in the PPS, in order to take into account the effects of all the oscillations in the angular power spectra, we calculate the 
angular power spectra at all $\ell$'s to avoid interpolation. Moreover, the accuracy that we use for our analyses, ensures the transfer function is well sampled in 
$k$-space that resolves the finest oscillations in the PPS in the convolution integration. We also make use of the unbinned Plik likelihood for high-$\ell$ TT and TTTEEE datasets
in order to obtain the best fits. However, we use the binned Plik likelihood for the MCMC analyses since we did not find noticeable differences in the best fit $\Delta\chi^2$
and the cosmological parameters and calculation of unbinned likelihood is significantly slower compared to the binned ones. 
For completeness too, since unbinned data for CMB polarization alone is not publicly available, in order to compare the individual constraints on the cosmological parameters, 
binned likelihood are more useful in our MCMC runs. 

We use BICEP2-Keck likelihood from the joint BICEP2/Keck and Planck analysis (BKP)~\cite{BICEP2Planck:2015joint}. In our analyses we use 5 bandpowers in the range $\ell=20-200$, 
from 150 GHz band of BICEP2/Keck and 217GHz and 353GHz bands of Planck. Needless to mention we compute and use the tensor power spectra from the models when we confront the models 
with BKP datasets.

We use the BINGO-2.0 version in order to evaluate the bispectra in equilateral and arbitrary triangular configurations. We have defined ${\rm M_{Pl}}^2=1/(8\pi G)=1$ and 
used $\hslash=c=1$ throughout the paper. 


\section{Results and discussions}\label{sec:results}

We present our results in this section in the following manner. We provide the best fit primordial power spectrum that we obtained from the complete
datasets. We tabulate the best likelihood values and the improvement in fit from WWI compared to the power law PPS. Using the best fit values,
we compare the best fit angular power spectra of temperature and polarization and their cross-correlations {\it w.r.t} best fit Planck baseline models.
We compare the background parameter constraints from the MCMC analyses against different datasets.
Finally, we present the bispectra in equilateral and arbitrary triangular configurations for the best fit models.

\subsection{Primordial scalar perturbation power spectrum}

\begin{figure*}[!htb]
\begin{center} 
\resizebox{142pt}{105pt}{\includegraphics{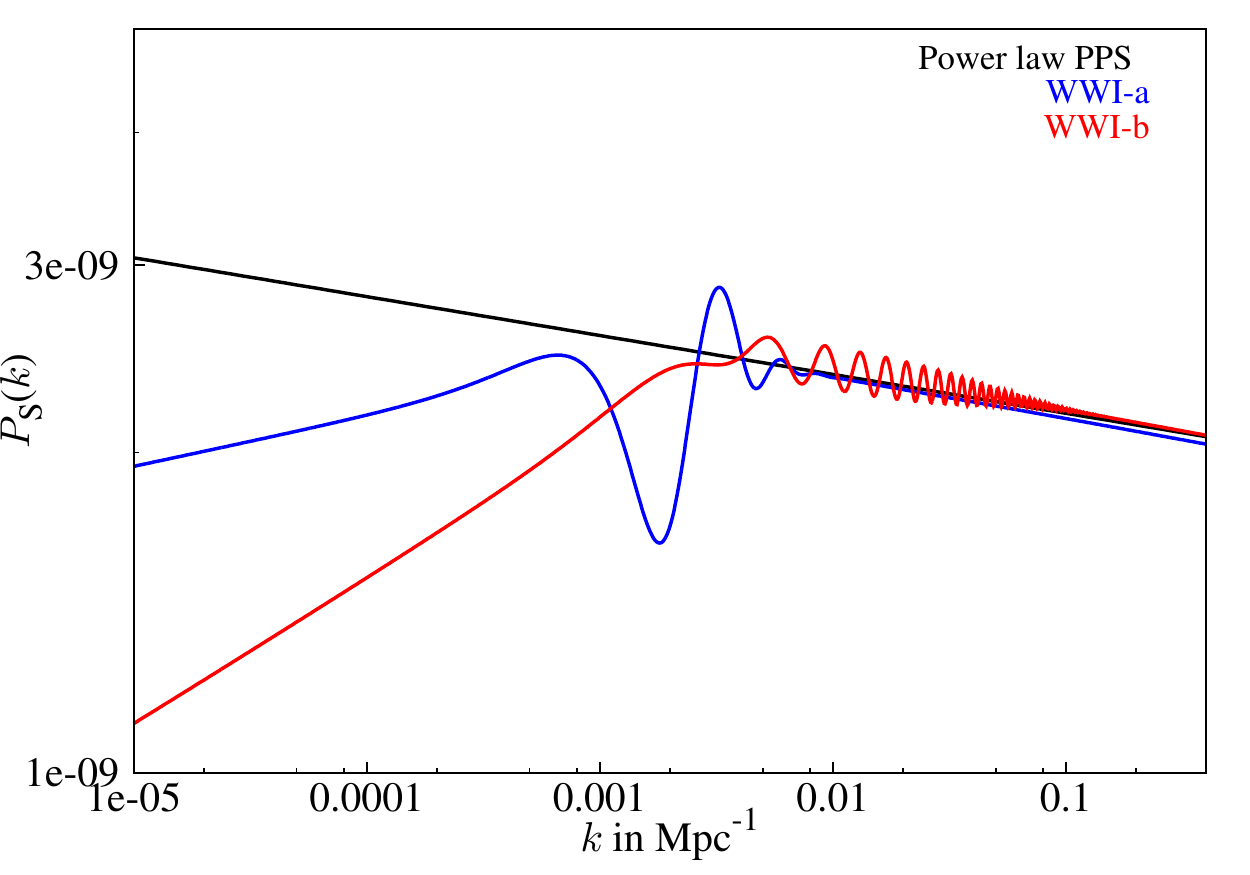}} 
\resizebox{142pt}{105pt}{\includegraphics{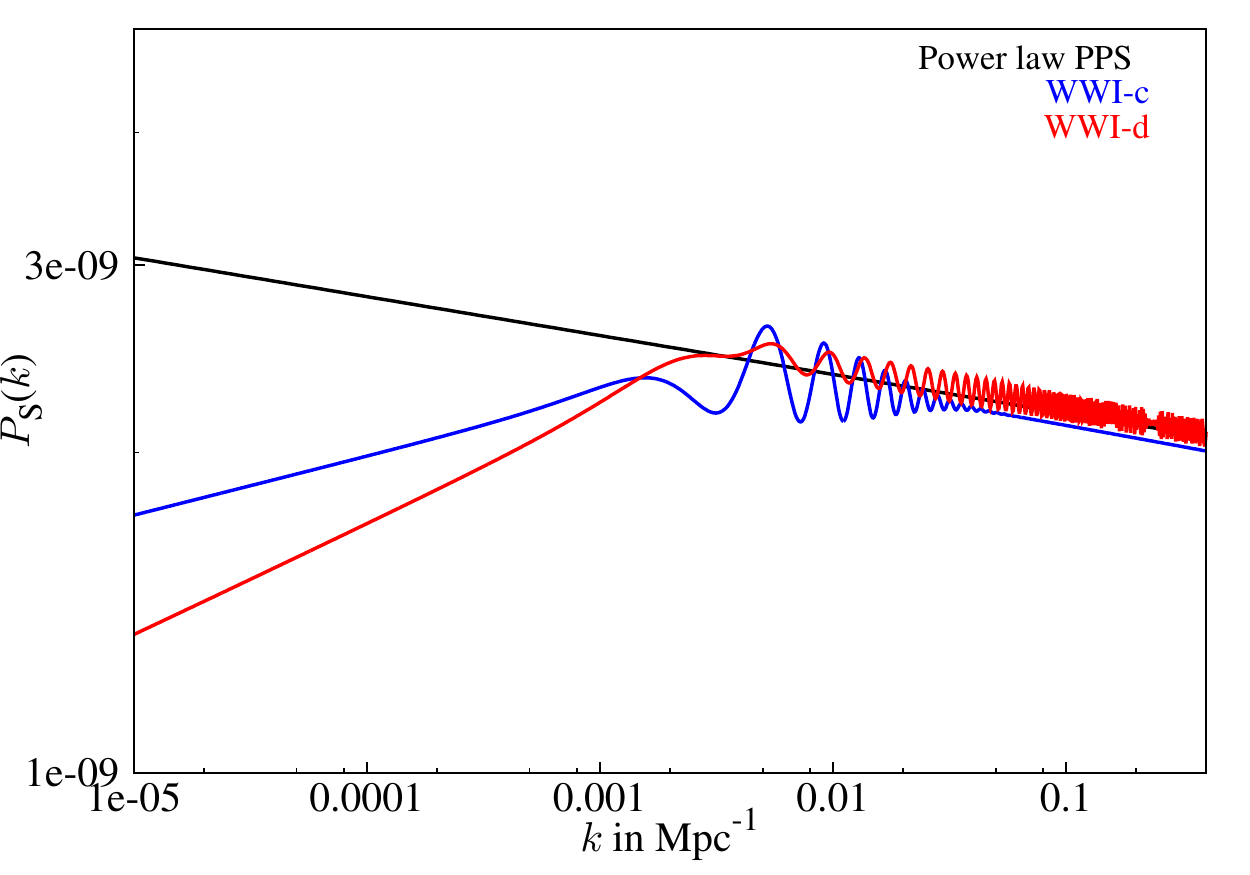}} 
\resizebox{142pt}{105pt}{\includegraphics{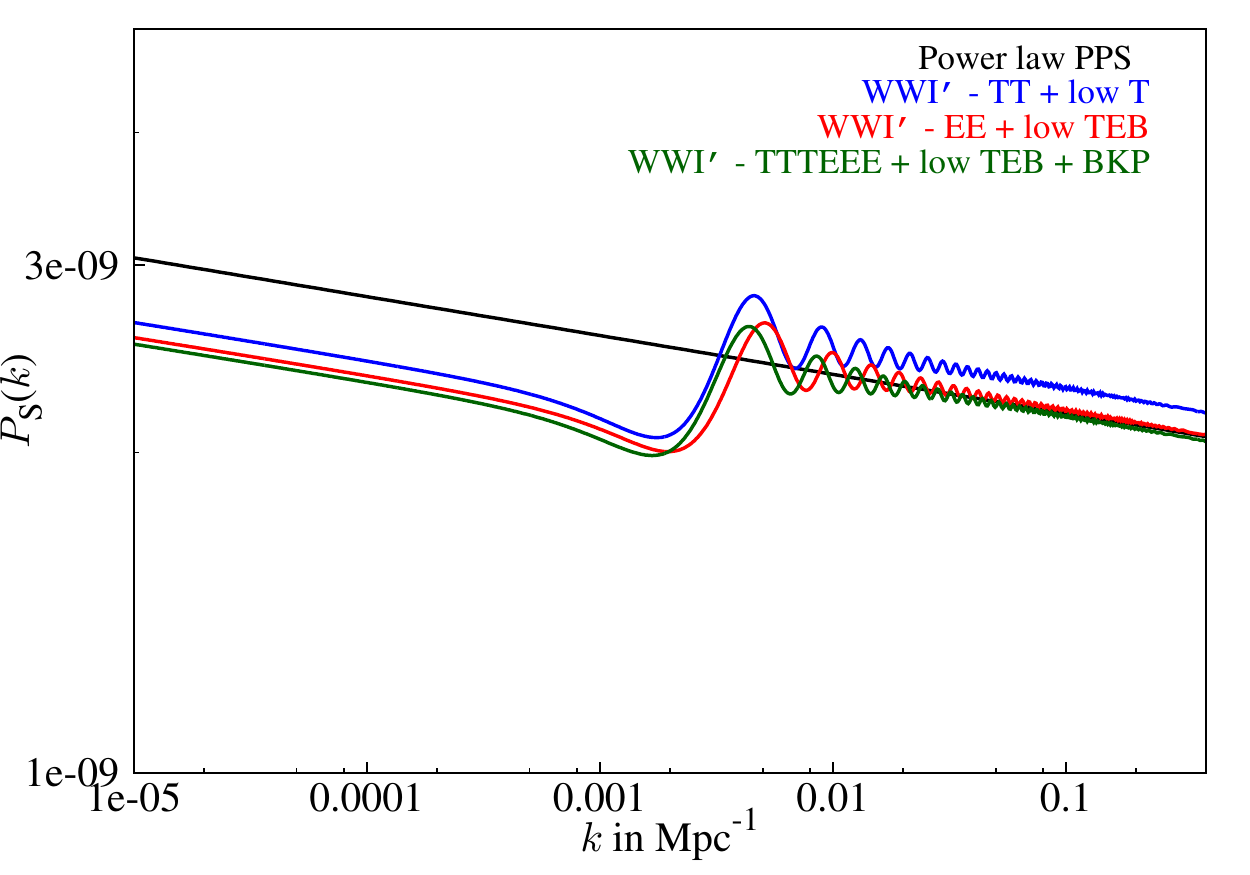}} 
\end{center}
\caption{\footnotesize\label{fig:psk} Wiggly Whipped Inflation : Best fit primordial power spectra. Left plot contains the local minima, WWI-[a,b] and the middle plot contains WWI-[c,d] respectively (Eq.~\ref{eq:equation-WWI}).
WWI-a provides $\sim7-8$ improvement in $\chi^2$ fit to temperature only data. Mainly the improvement comes from the large scale power suppression and $\ell\sim22-40$ region.
WWI-c provides $\sim10$ improvement in $\chi^2$ fit to polarization data. This improvement comes from low-$\ell$ TEB and high-$\ell$ E data.
WWI-b provides $\sim11$ and WWI-d provides more than $13$ improvement to combined temperature and polarization datasets. In this case, most of the improvement comes owing to the inability of 
baseline model in fitting the temperature and polarization datasets in a combination. The plot at the right contains the WWI$'$ (Eq.~\ref{eq:equation-WWI'}) best fit when compared with T, E and combined datasets. For the combined 
datasets, WWI$'$ provides nearly 12 improvement in the fit compared to power law best fit.}
\end{figure*}

We have demonstrated that WWI offers a wide variety of features in the primordial power spectra. When compared against the 
Planck datasets we locate 4 local minima that provide substantial better fit compared to the power law power spectra. We categorize the local regions 
of parameter space by their distinct nature of features. We denote them as WWI-[a,b,c,d]. We plot these models in left and middle plots of Fig.~\ref{fig:psk}. Here we should 
stress that WWI-[a,b,c,d] that are plotted in the figure, in strict sense, are the local minima of the TTTEEE + lowTEB + BKP combination. WWI-a is actually the global 
minima of TT + lowT dataset, that turns out to be a local minima of the complete datasets. WWI-c on the other hand, has characteristic wide scale oscillations that provide
improvement in fit to the high-$\ell$ EE data. We search for the local and global minima in other individual and combination of datasets in the vicinity of WWI-[a,b,c,d]. Obviously the best fit 
parameters will be different when we compare different datasets. However, the broad shape of that particular feature remains similar in all the datasets since 
the WWI potential parameters do not change significantly. The right plot of the same figure provide best fit PPS from WWI$'$ when compared
against T, E and combined datasets.

\subsection{Best fit results}

In Table~\ref{tab:bestfits} we tabulate the best fit $-2\log{\rm [Likelihood]}$ for the WWI-[a,b,c,d], WWI$'$ and the Planck baseline model. 
Each row block of the table contains the data combination that we used. From top to bottom we provide the analyses for TT + lowT, 
EE + lowTEB, TTTEEE + lowTEB, TT + TE + EE + lowTEB + BICEP-Keck-Planck dust, TTTEEE + lowTEB + BICEP-Keck-Planck dust, unbinned TT + lowT and unbinned TTTEEE + lowTEB. Here we should 
mention again that the cosmological parameters do change when compared with different combination of datasets. For example WWI-a that is 
a local minima of TTTEEE + lowTEB (as plotted in Fig.~\ref{fig:psk}) is not {\it strictly} a local minima of TT + lowT (the position and the amplitude 
of the features may vary a little). We find with a marginal change in parameter values, a PPS very close in shape to WWI-a, represent the global 
best fit to TT + lowT datasets. Since Powell's minimizer algorithm converges to a point very close to the parameter space of the starting point
in our cases, we searched for the local best fit to TTTEEE + lowTEB + BKP data in the vicinity of the global best fit to TT + lowT data. 
Hence, WWI-[a,b,c,d] represent a class of resembling PPS that can possibly be a global best fit to a subset of data but in strict sense they 
are local best fit to the TTTEEE + lowTEB + BKP dataset. Obtained potential parameters, $\l[\ln (10^{10} V_i),~\phi_{01},~\gamma,~\phi_{\rm T},~\ln(\Delta)\r]$ for TTTEEE + lowTEB + BKP 
are $\l[1.73,~0.099,~0.019,~7.88,~-4.47~\r]$ (for WWI-a), $\l[1.75,~0.019,~0.04,~7.91,-7.07~\r]$ (for WWI-b), 
$\l[1.72,~0.039,~0.02,~7.91,-6.04~\r]$ (for WWI-c) and $\l[1.76,~0.017,~0.03,~7.91,-11~\r]$ (for WWI-d). However, since WWI$'$ has a unique shaped PPS, in Fig.~\ref{fig:psk} we plotted the obtained global best fits of 
TT + lowT, EE + lowTEB and TTTEEE + lowTEB + BKP datasets. For WWI$'$, the obtained values for inflaton potential parameters, $\l[\ln (10^{10} V_i),~\phi_{01},~\phi_{\rm T}\r]$, are 
$\l[0.344,~0.12,~4.5\r]$ (for TT + lowT), $\l[0.295,~0.11,~4.5\r]$ (for EE + lowTEB) and $\l[0.282,~0.11,~4.51\r]$ (for TTTEEE + lowTEB + BKP).

\renewcommand{\arraystretch}{1.1}
\begin{table*}[!htb]
\footnotesize{
\begin{center}
\vspace{4pt}
\begin{tabular}{| c | c | c | c | c | c | c |}
\hline\hline
\multicolumn{7}{|c|}{\bf Individual likelihoods comparison}\\

\hline
 Individual & Baseline & WWI-a & WWI-b & WWI-c & WWI-d & WWI$'$\\
 likelihood &  & $\Delta_{\rm DOF}=4$&$\Delta_{\rm DOF}=4$ & $\Delta_{\rm DOF}=4$&$\Delta_{\rm DOF}=4$ &$\Delta_{\rm DOF}=2$\\
\hline
 TT	&761.1 	&762	&761.9	&762.8	&762.8 	& 762.4\\
 lowT 	&15.4	&8.2 	& 13.4 	&12.1 	&13 	& 10.2\\
 Total 	&778.1	&772.1 (-6) 	&777 (-1.1)  	&777 (-1.1) 	&778.4 (0.3)	& 775 (-3.1)\\

\hline
 EE	&751.2 	& 748.8	& 747.2	 & 748.6 &750.2   & 746.8\\
 lowTEB &10493.6&10490 	&10495.6 &10492.4&10495.7 &10492.2\\
 Total 	&11248.8&11241.8 (-7)&11246.2 (-2.6) &11244.5 (-4.3)&11249.3 (0.5) &11242.3 (-6.5) \\
 
 \hline
 TTTEEE &2431.7 &2432.7 & 2422.6  & 2427.8&2421.7  &2426.5\\
 lowTEB &10497 	&10490.8& 10495.1 &10493.4&10495.3 &10492.7\\
 Total 	&12935.6&12929.5 (-6.1)& 12924.2 (-11.4) &12927.6 (-8)&12923.4 (-12.2) &12925.2 (-10.4) \\

 \hline
 TT 	& 764.5	 &763.6	 &762.2	  &764.4  &762.9  &762.8 \\
 EE 	& 753.9	 &754.8	 &750.5	  &750.8  &750.8  &751\\
 TE 	& 932	 &933.4	 &928.7	  &929.2  &927	  &928.8\\ 
 lowTEB & 10498.4&10490.4&10495.8 &10493.7&10495.6&10492.4\\
 BKP 	& 41.6	 &42	 &42	  &42.6   &41.8   &42.9\\
 Total 	&12997 	 &12991 (-6)  &12985.9 (-11.1) &12987.2 (-9.8)&12985 (-12)   &12985.1 (-11.9) \\

\hline
 TTTEEE & 2431.7&2432.8	 &2421.4 &2426.7  &2421    &2425.7 \\
 lowTEB &10498.5&10490.5 &10495.5&10493.6 &10495.8 &10492.6\\
 BKP 	& 41.6	&42	 &42.7	 &42	  &41.9    &42.5	\\
 Total	&12978.3&12971.3 (-7) &12967.3 (-11)&12968.6 (-9.7) &12965 (-13.3)   &12968.6 (-9.7) \\
\hline
 TT (bin1) 	&8402.1&8404.1	 &8403.9 & 8405.2 &8402.1   &8401.9 \\
 lowT		&15.4  &8.3 	 &13.3   & 11.9   &13.2     &10.3\\
 Total		&8419.6&8414.7 (-4.9)&8419.5 (-0.1) & 8419.8 (0.2) &8418.1 (-1.5)   &8414.4 (-5.2) \\
\hline
 TTTEEE (bin1) &24158.2&24158.6 &24149   &24155  &24148.4 & 24151.5\\
 lowTEB 	   &10497.6&10490.3 &10493.4 &10493.6 &10495.3 & 10492.7\\
 Total		   &34661.9&34655.3 (-6.6)&34650.5 (-11.4)&34654.4 (-7.5) &34649.5 (-12.4)& 34650.6 (-11.3) \\
 
\hline\hline
\end{tabular}
\end{center}
\caption{~\label{tab:bestfits} Best fit parameters for the Wiggly Whipped Inflaton potential. We provide the best fit 
$-2 \ln [{\rm Likelihood}]$ for individual datasets for the types of WWI features denoted as WWI-[a,b,c,d], WWI$'$ and that are plotted
in Fig~\ref{fig:psk}. Note that while WWI-[a,c] provide local best fits to individual and combined data, WWI-[b,d] seem to capture 
the global best fit by providing improvement in all datasets (TT, TE, EE) in a combination when compared with power law model.
The best fit background, WWI potential and nuisance parameters do indeed change to certain extent in different dataset combinations 
but the primordial power spectrum remains very similar to the ones plotted in Fig.~\ref{fig:psk} for all dataset combinations.
In case of WWI$'$, note that the similar shaped features (shown in right plot of Fig.~\ref{fig:psk}) are able to address both temperature and polarization data 
individually and in combinations. We find similar improvement in fit ($\Delta\chi^2\sim-12$ compared to power law) as in WWI-[b,d] but here only with 
2 extra parameters. WWI$'$ also provides the best likelihood to the high -$\ell$ EE data. We do not provide in the table, the ${\cal O}(1)$ improvement in fits 
from the priors in the nuisance and calibration parameters. $\Delta_{\rm DOF}$  indicates extra parameters used in the models compared to its featureless potential 
parameters. The `bin1' in the last 2 row blocks indicate unbinned high-$\ell$ data have been used to obtain the best fits. In the `Total' row we also 
provide the differences in the $\chi^2$ {\it w.r.t} the power law model within brackets.}} 
\end{table*}

\begin{figure*}[!htb]
\begin{center} 
\resizebox{420pt}{320pt}{\includegraphics{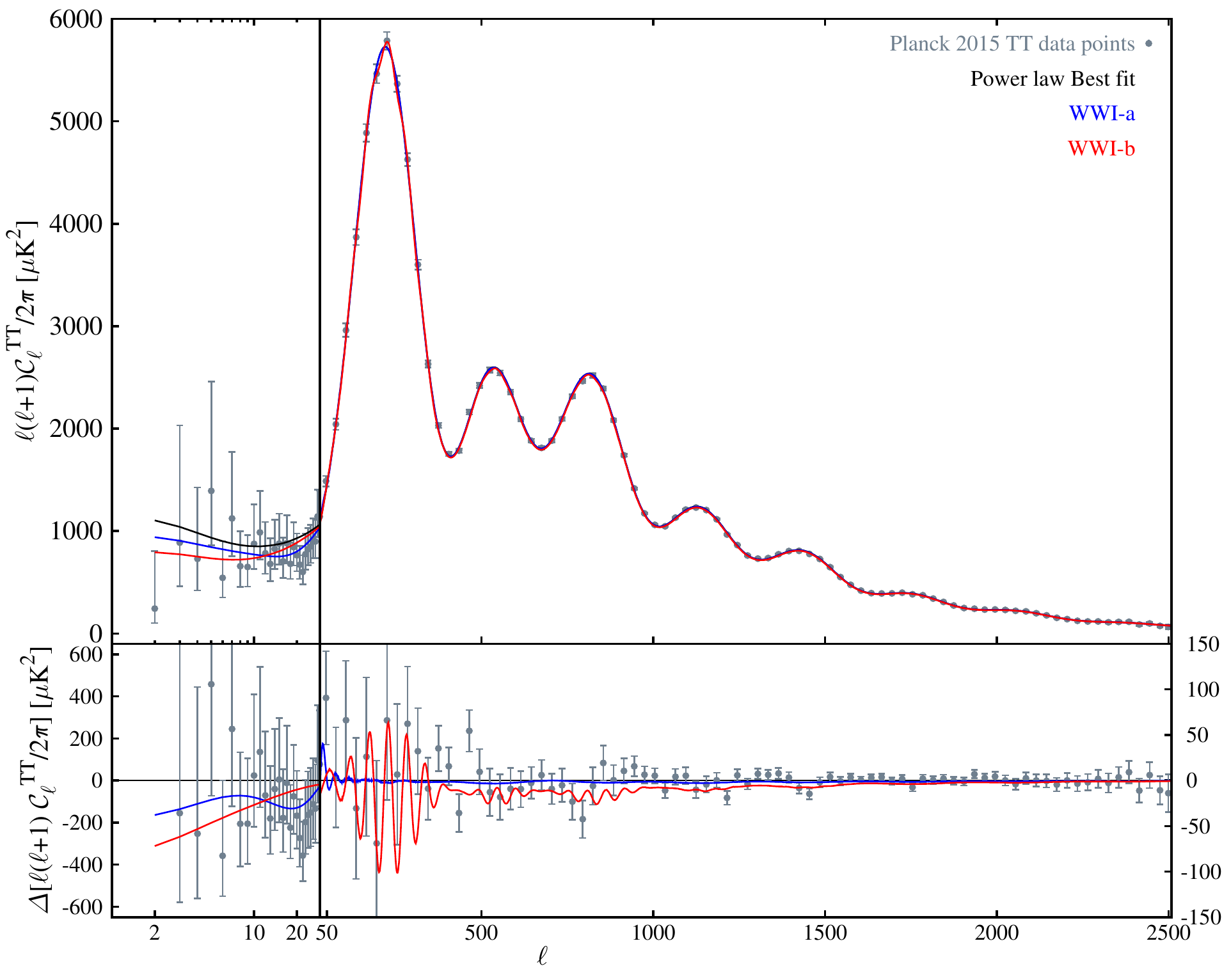}} 
\resizebox{210pt}{160pt}{\includegraphics{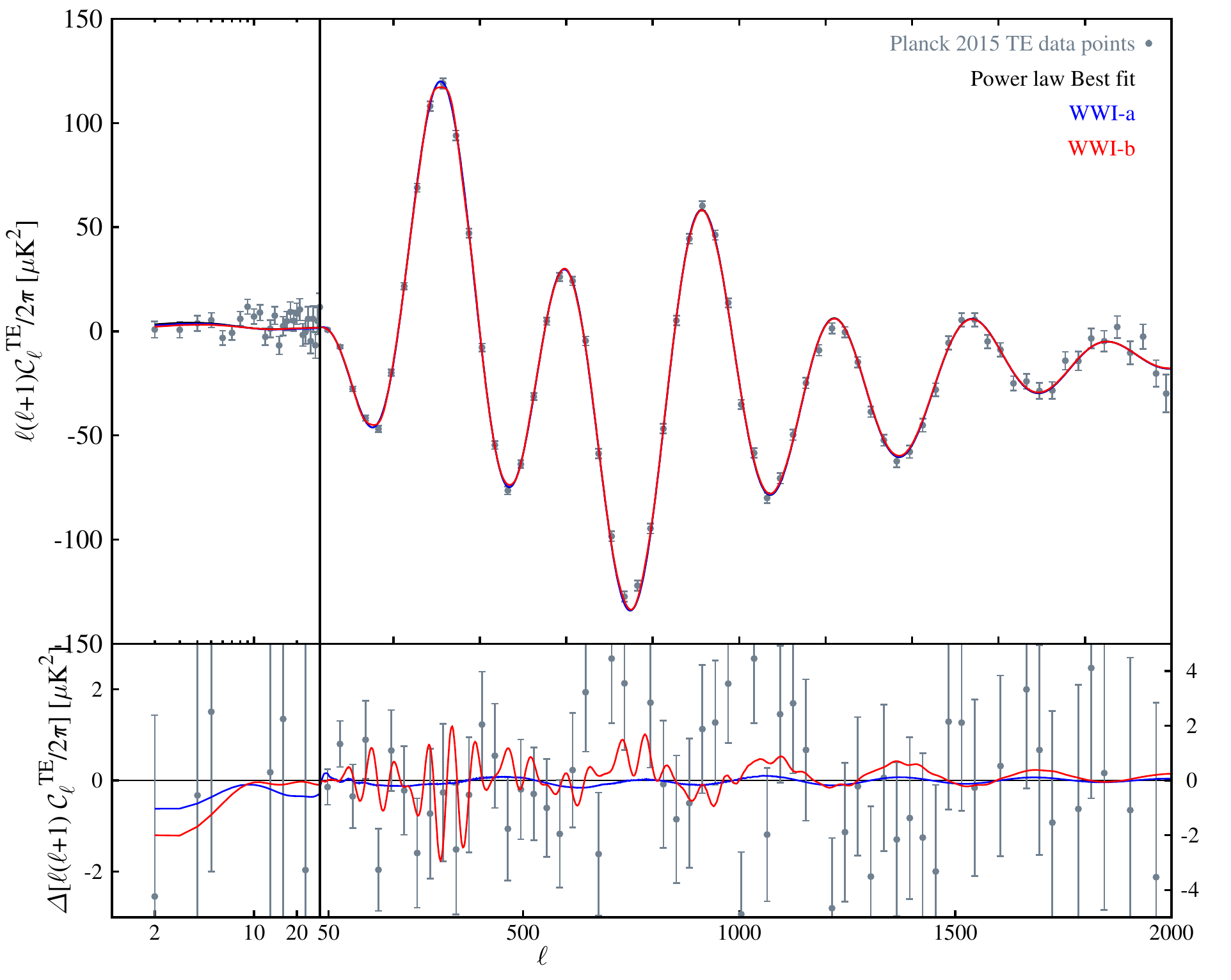}} 
\resizebox{210pt}{160pt}{\includegraphics{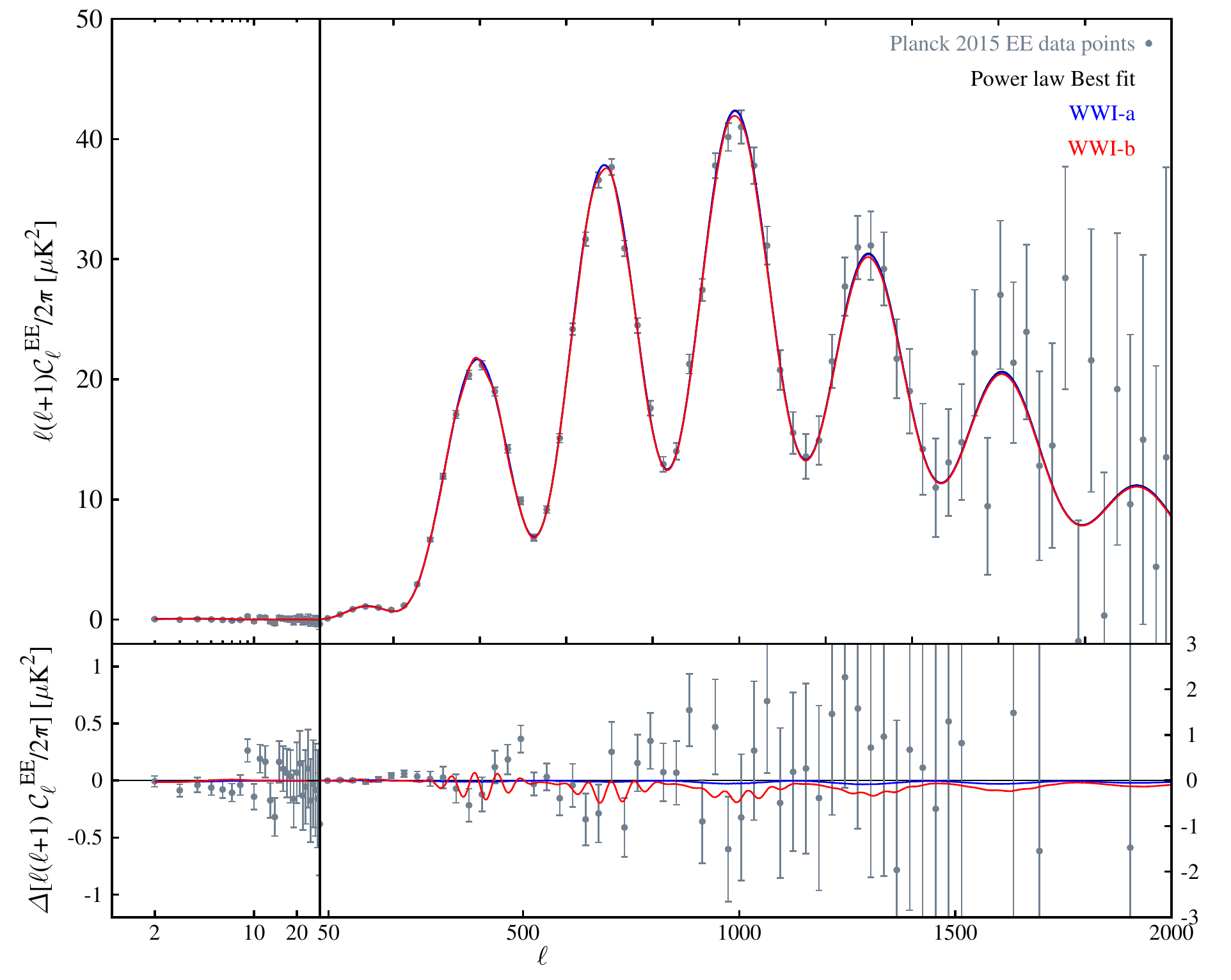}} 
\end{center}
\caption{\footnotesize\label{fig:clab} Wiggly Whipped Inflation : Angular power spectra for temperature and polarization anisotropies. Theoretical 
predictions from WWI-a and WWI-b are provided. These are the best fits to TTTEEE + lowTEB + BKP datasets. Power law baseline best fit is provided in 
black. Power spectra from WWI and the data, that residual from power law best fit model are plotted below to highlight the features in the angular power spectra.}
\end{figure*}

\begin{figure*}[!htb]
\begin{center} 
\resizebox{420pt}{320pt}{\includegraphics{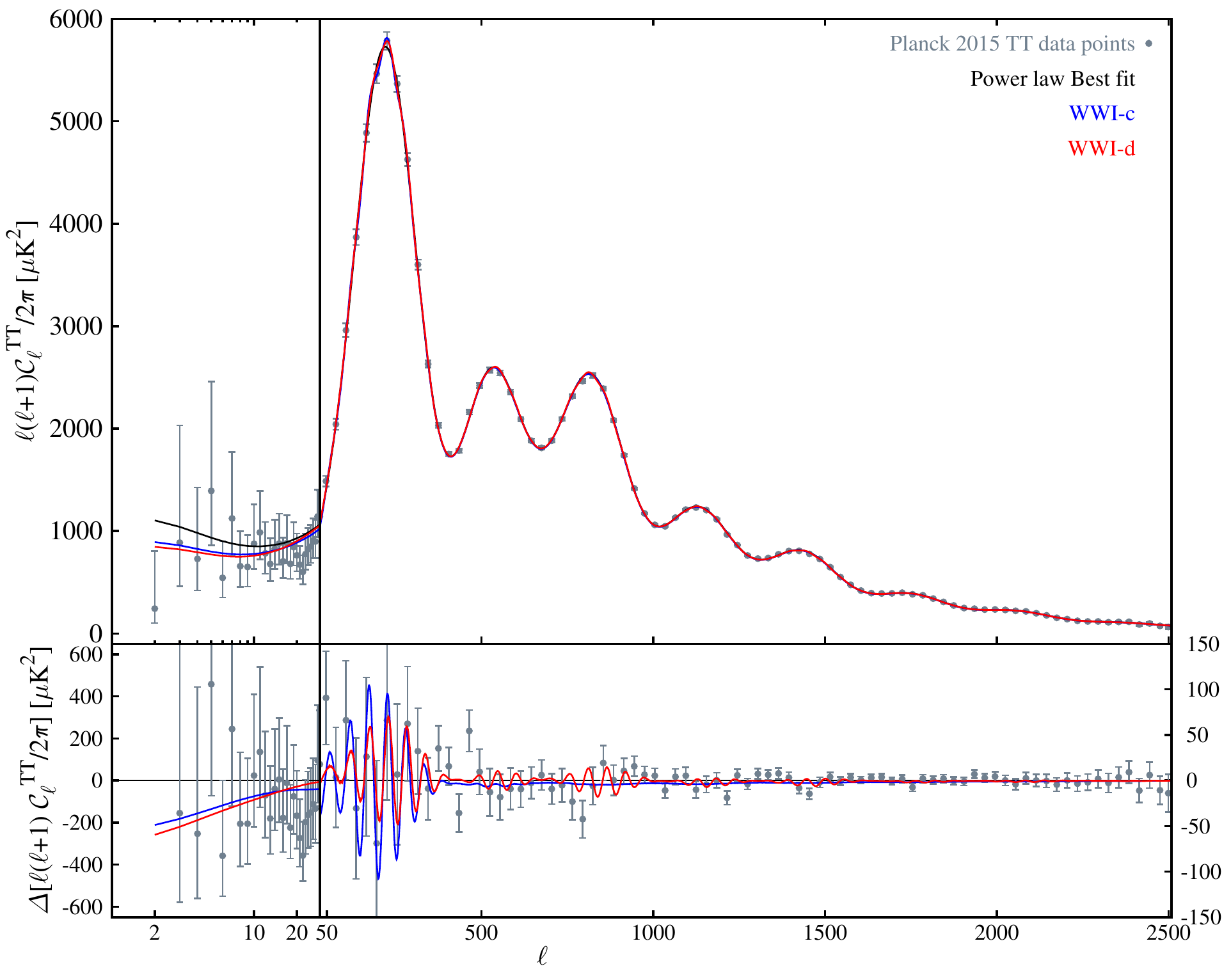}} 
\resizebox{210pt}{160pt}{\includegraphics{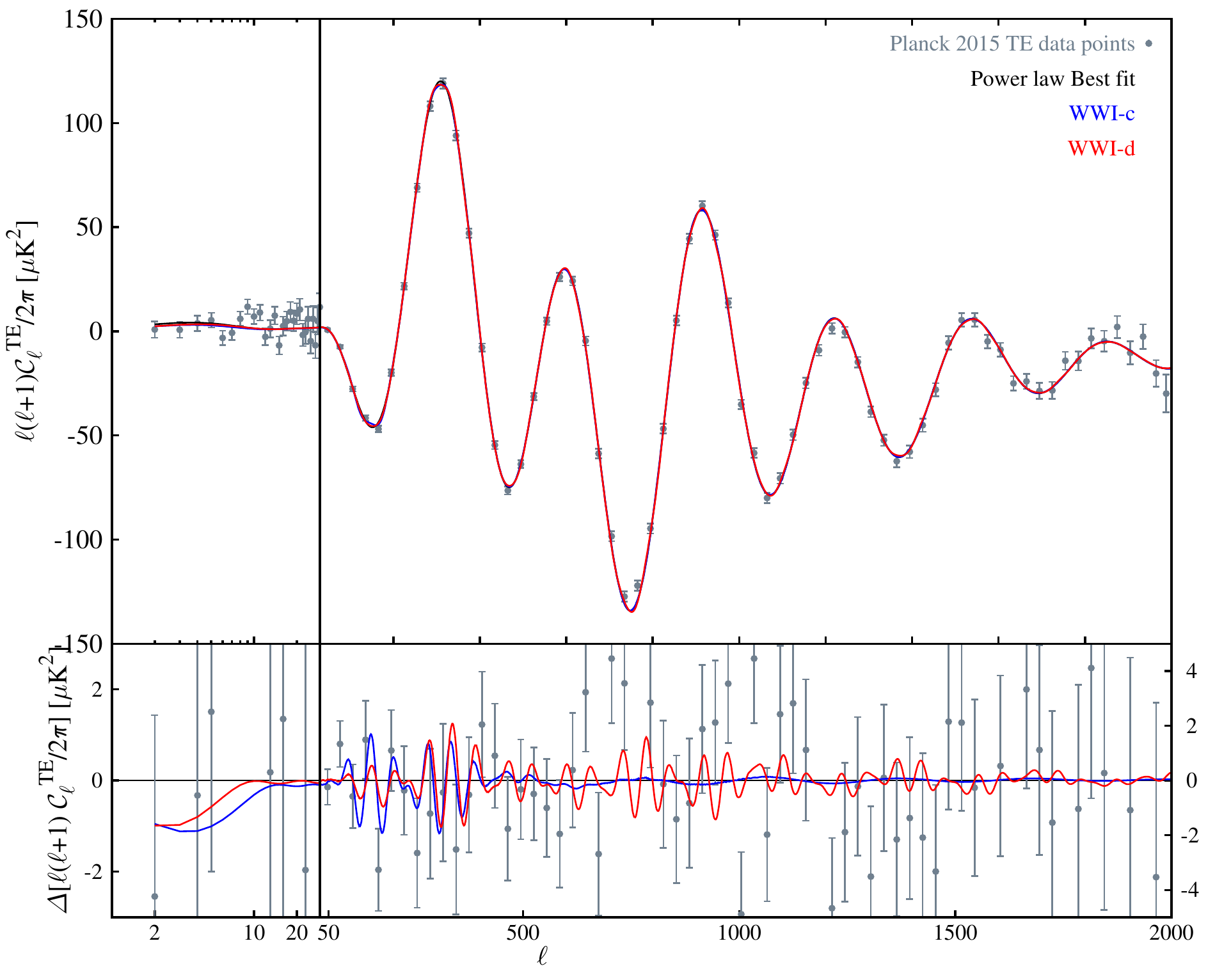}} 
\resizebox{210pt}{160pt}{\includegraphics{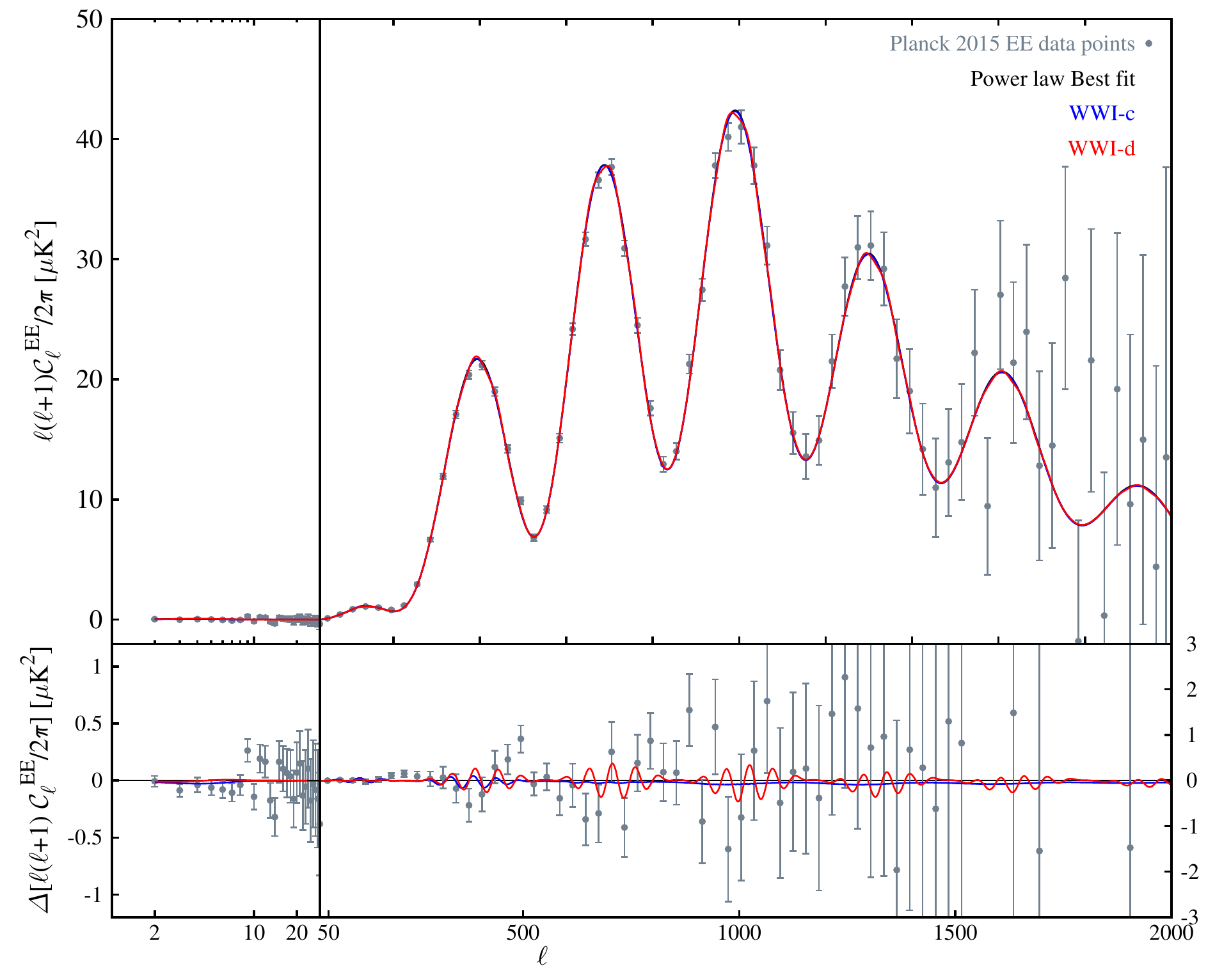}} 
\end{center}
\caption{\footnotesize\label{fig:clcd} Wiggly Whipped Inflation : Angular power spectra for temperature and polarization anisotropies. Theoretical 
predictions from WWI-c and WWI-d are provided. These are the best fits to TTTEEE + lowTEB + BKP datasets. Power law baseline best fit is provided in 
black. Power spectra from WWI and the data, that are residual from power law best fit model are plotted below to highlight the features in the angular power spectra.}
\end{figure*}

\begin{figure*}[!htb]
\begin{center} 
\resizebox{420pt}{320pt}{\includegraphics{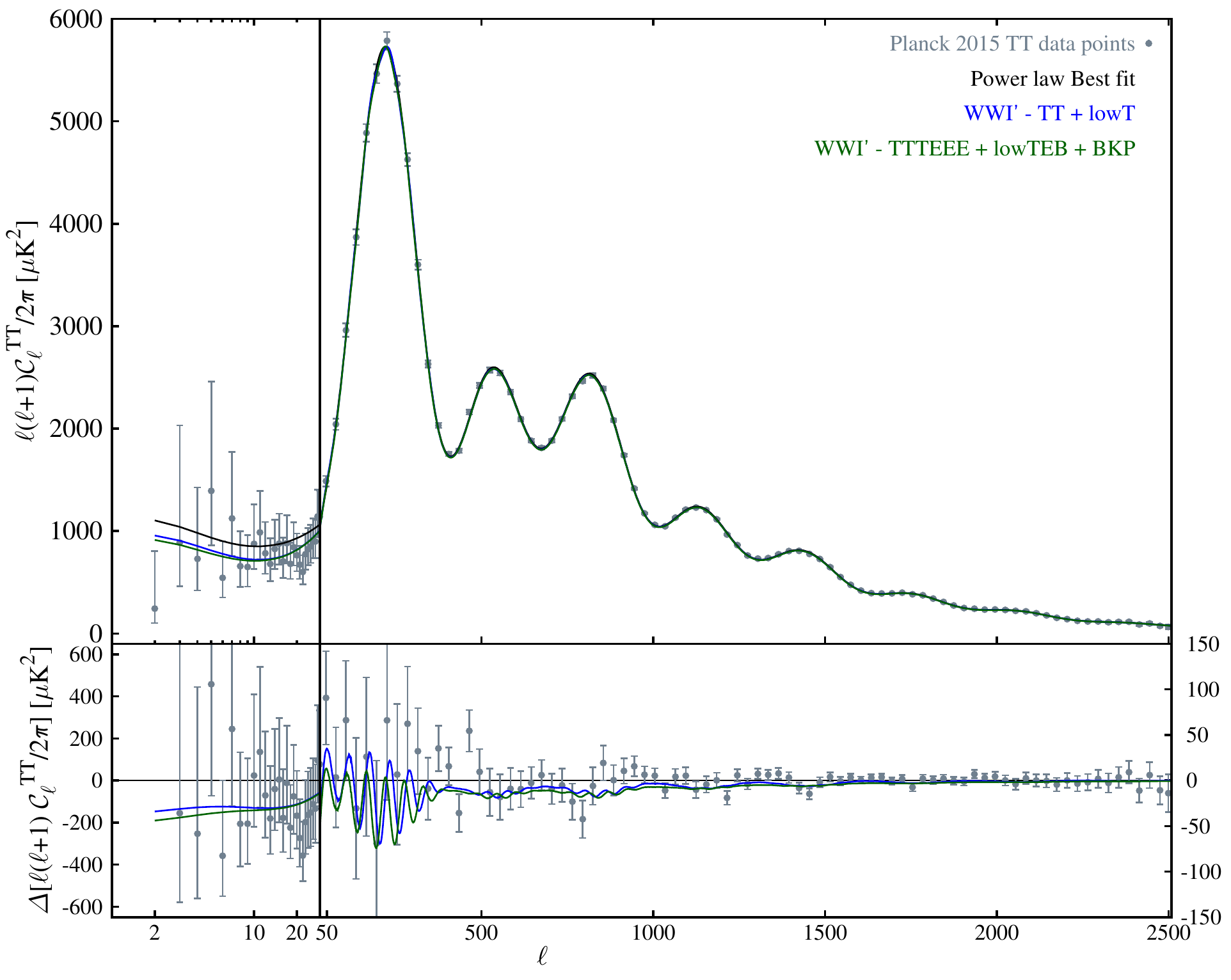}} 
\resizebox{210pt}{160pt}{\includegraphics{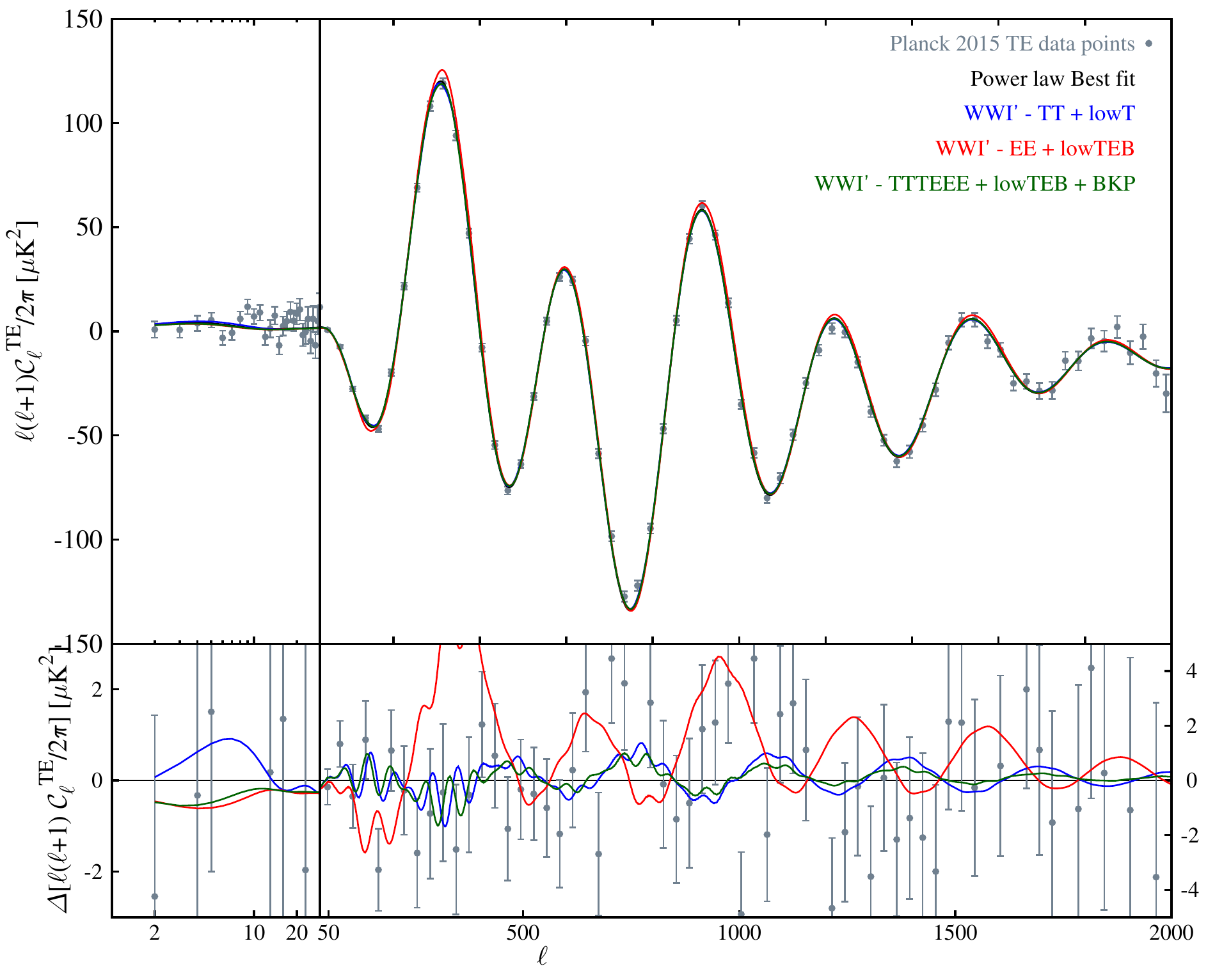}} 
\resizebox{210pt}{160pt}{\includegraphics{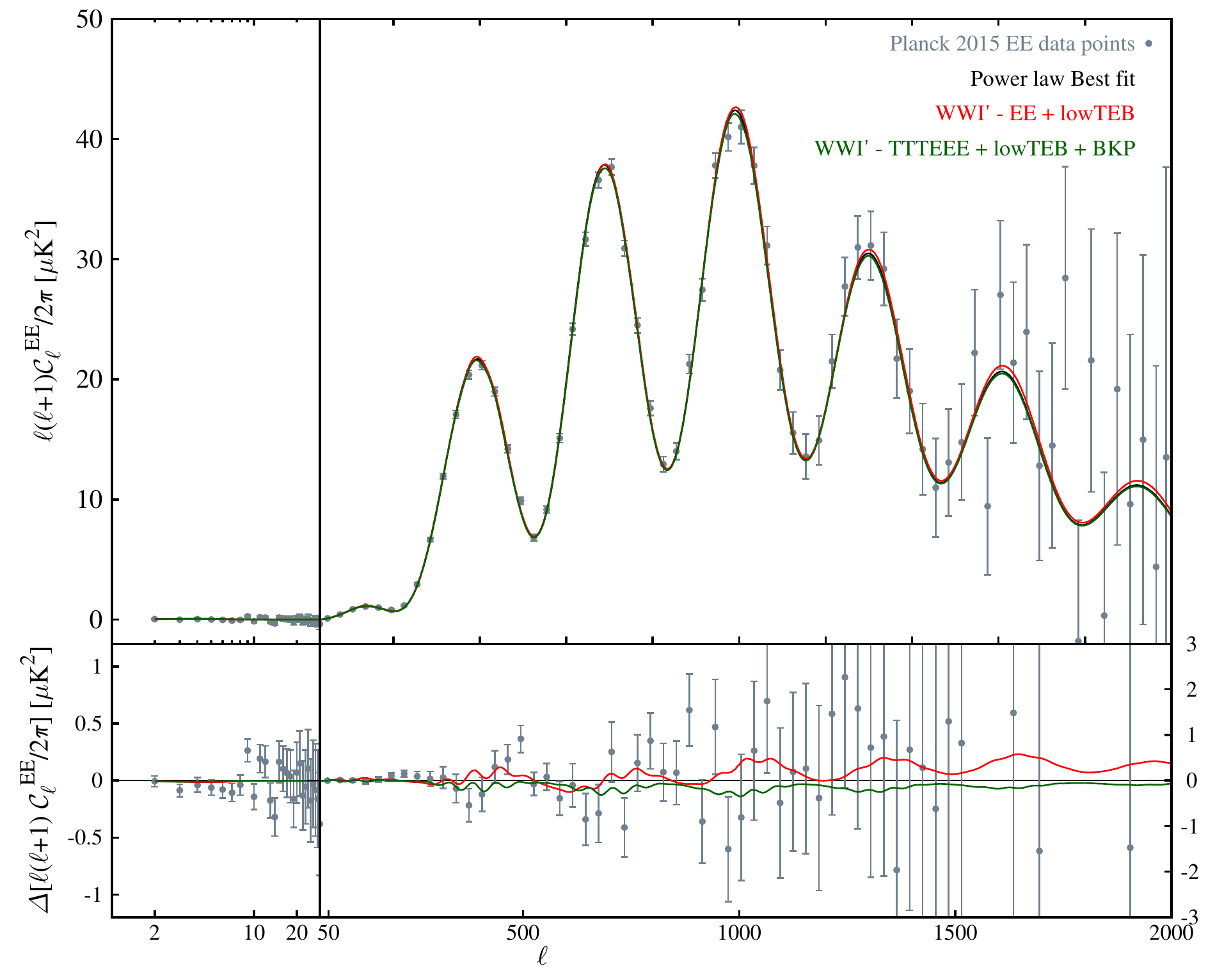}} 
\end{center}
\caption{\footnotesize\label{fig:clprime} Wiggly Whipped Inflation : Angular power spectra for temperature and polarization anisotropies. Theoretical 
predictions from WWI$'$ are provided. These are the best fits to TT + lowT, EE+lowTEB and TTTEEE + lowTEB + BKP datasets. Power law baseline best fit is provided in 
black. Power spectra from WWI$'$ and the data, that are residual from power law best fit model are plotted below to highlight the features in the angular power spectra.}
\end{figure*}

Note that WWI-a provides improvement in likelihood principally for the lowT case. The localized feature similar to step in the inflaton potential
around $k\sim0.002~{\rm Mpc}^{-1}$ ($\ell\sim22$) along with the large scale power suppression provides around 8 improvement in fit to the lowT and lowTEB data.
WWI-a also provide some improvement to the high-$\ell$ EE data. WWI-c which introduces wiggles in the PPS ranging $k\sim0.002-0.05~{\rm Mpc}^{-1}$
provide a moderate improvement to lowT and lowTEB datasets due to the large scale suppression but fails to capture the $\ell\sim22$ dip in TT. However
the wide wiggles provide a better fit to EE and TE-datasets. WWI-b and WWI-d, though having wiggles in the primordial power spectrum, do not provide 
better fit to individual temperature or polarization data at high-$\ell$ but they provide an overall better fit when the complete datasets are compared with. 
Note that when we combine T and E datasets (TT + TE + EE + lowTEB + BICEP-Keck-Planck dust), the baseline model provide worse fit to TT and EE datasets 
{\it w.r.t.} their values when compared individually (TT + lowT and EE + lowTEB). This difference arise because temperature data favors a low baryon
density ($\Omega_{\rm b}{h^2}\sim 0.0222$) while E-polarization goes for a higher ($\Omega_{\rm b}{h^2}\sim 0.024$) value. The baseline model can not trade-off for the baryon
density and hence settles for marginal worse likelihood to all datasets. We carry out the analysis with TT + TE + EE + lowTEB + BICEP-Keck-Planck dust specifically 
to point out the individual worse fits. WWI-b and WWI-d manage to compensate for the baseline worse fit by providing wiggles in the PPS that can address EE data better 
than the baseline model {\it but} at the same time keeping a low baryon density ($\Omega_{\rm b}{h^2}\sim 0.023$). These two spectra represent the global fit to the overall Planck 
data. Though the WWI-b and WWI-d are indistinguishable from the likelihood values, they have distinct signatures in the power spectrum. The Wiggles in WWI-b damps down at 
$k\sim0.1~{\rm Mpc}^{-1}$ while WWI-d continues to be present at smaller scales. The difference between these two spectra becomes evident when we calculate the three point 
correlations {\it i.e.} the bispectra. It is interesting that WWI$'$, with a fixed shaped power spectra feature, is able to provide better fit to the temperature and polarization
data individually and also in combinations. In Fig.~\ref{fig:psk} (the right plot), we have provided the best fit PPS for WWI corresponding to the best fit provided in Table~\ref{tab:bestfits}.
We find $\Delta\chi^2\sim-5$ and $-6$ {\it w.r.t.} power law when we compare WWI$'$ with the temperature and polarization data respectively. This particular best fit do not provide 
better fit to the high-$\ell$ TT data but the best fit for EE quoted in the Table is in very good agreement with the high-$\ell$ EE data. However, the position and amplitude of the 
oscillations required to have a good agreement with TT and EE data are different. When we compare WWI$'$ with the combined datasets, 
the best fit provides $\Delta\chi^2\sim -12$ compared to power law with only 2 extra parameters. Breakdown of likelihood in temperature and polarization data is very similar to 
WWI-[b,d] case, and we find, WWI$'$ with the PPS plotted in Fig.~\ref{fig:psk} is providing improvement to T and E datasets combinations from Planck since the best fit likelihoods 
from power law are degraded from the individual best fit values when temperature and polarization data are used separately.

In Fig.~\ref{fig:clab} and~\ref{fig:clcd}, we plot the best fit angular power spectra for temperature and polarization anisotropies from WWI and 
the Planck data. In both the plots we have also plotted the best fit results from the power law PPS in black. In each plot, the bottom panel 
represent the data and the power spectra residual to the power law best fit. The left panel in each plot captures the multipoles $\ell=2-29$
and are plotted in log scales while the right panels display high-$\ell$ ($\ell=30-2508$ in case of TT and $\ell=30-1996$ in case of EE and TE) data 
and best fit results in linear scale. Note that, here the plotted results correspond to the best fit obtained against TTTEEE + lowTEB + BKP datasets.
Improvement in fit from the low-$\ell$ data is evident the suppression in the low-$\ell$ residual plot in all the best fits. Only WWI-a is able to 
address the dip around $\ell=22$ in a convincing manner. WWI-[b,c,d] introduce features at the high-$\ell$ as well. However, note that, at high-$\ell$
we do not get notable improvement in fit when we use temperature and polarization data separately, but in a joint analysis, WWI-b and WWI-d 
interestingly provide a noticeable better fit. In Fig.~\ref{fig:clprime} we plot the best fit angular power spectra and their residuals from power law best fit 
for WWI$'$. Unlike WWI model, in this plot we plot best fit from TT + lowT, EE + lowTEB and TTTEEE + lowTEB + BKP. For the TT plot we show best fit 
from TT + lowT and TTTEEE + lowTEB + BKP and for EE plot we present EE + lowTEB and TTTEEE + lowTEB + BKP best fits. For TE, we provide all three best fit
power spectra. In the TE plot, the EE best fit is seen to be not fitting the data well at the acoustic peaks. Since high-$\ell$ temperature data is not used in obtaining 
EE + lowTEB best fit, we can expect some disagreement between temperature and polarization best fits. The large mismatch here might point out a systematic tension between T and E 
data. However, we should note that the statistical uncertainties in the EE data is substantially larger than the TT data and differences in their best fits 
can just be an artifact of statistical fluctuations.

\subsection{Change in the parameter constraints}

\begin{figure*}[!htb]
\begin{center} 
\resizebox{142pt}{105pt}{\includegraphics{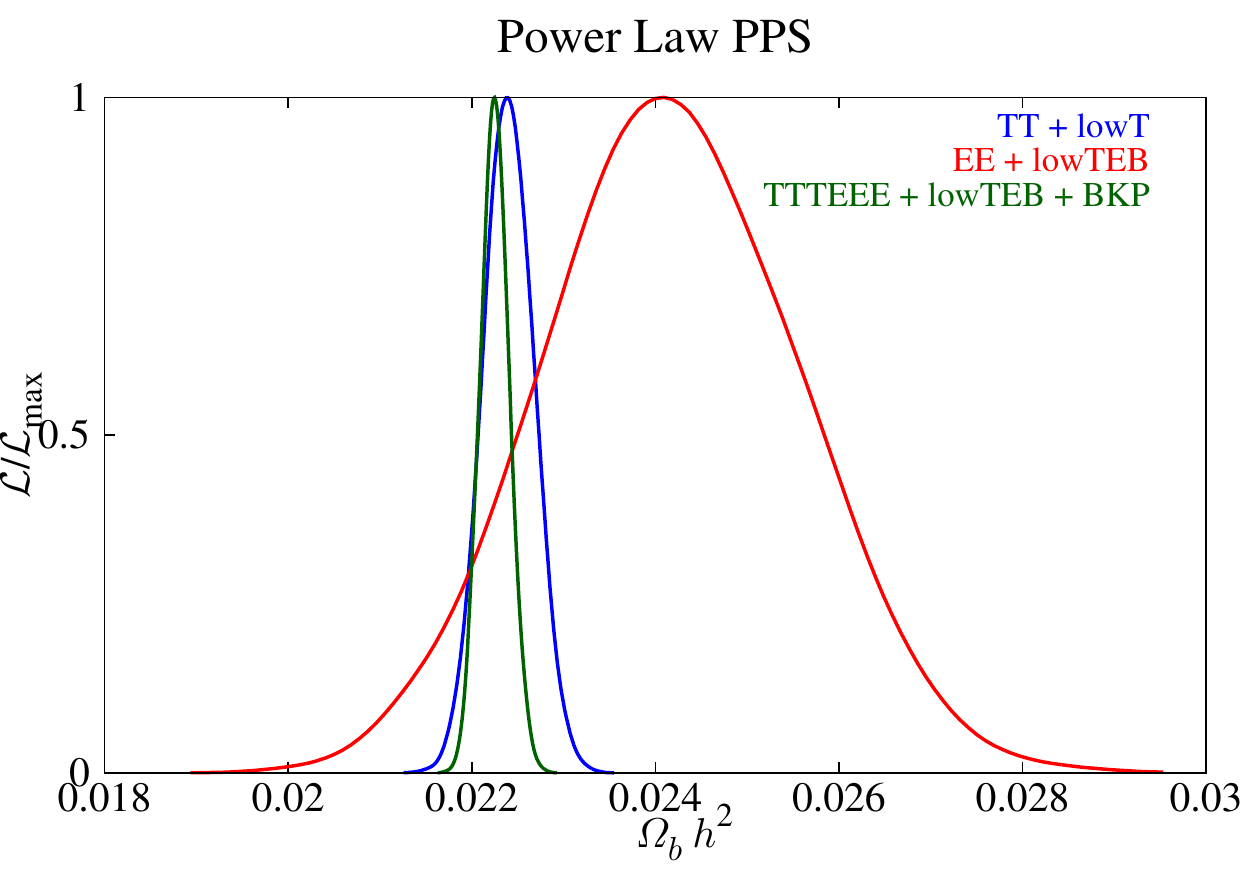}} 
\resizebox{142pt}{105pt}{\includegraphics{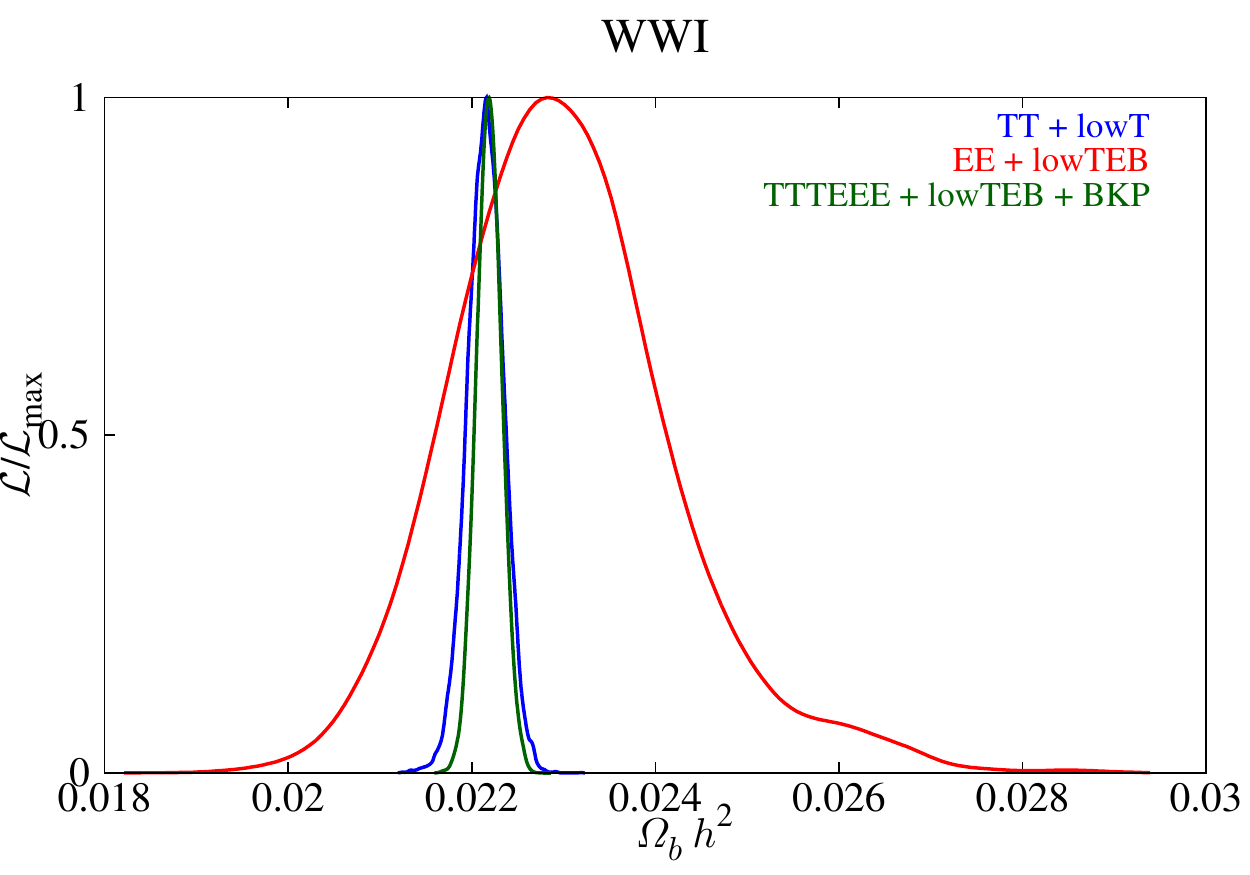}} 
\resizebox{142pt}{105pt}{\includegraphics{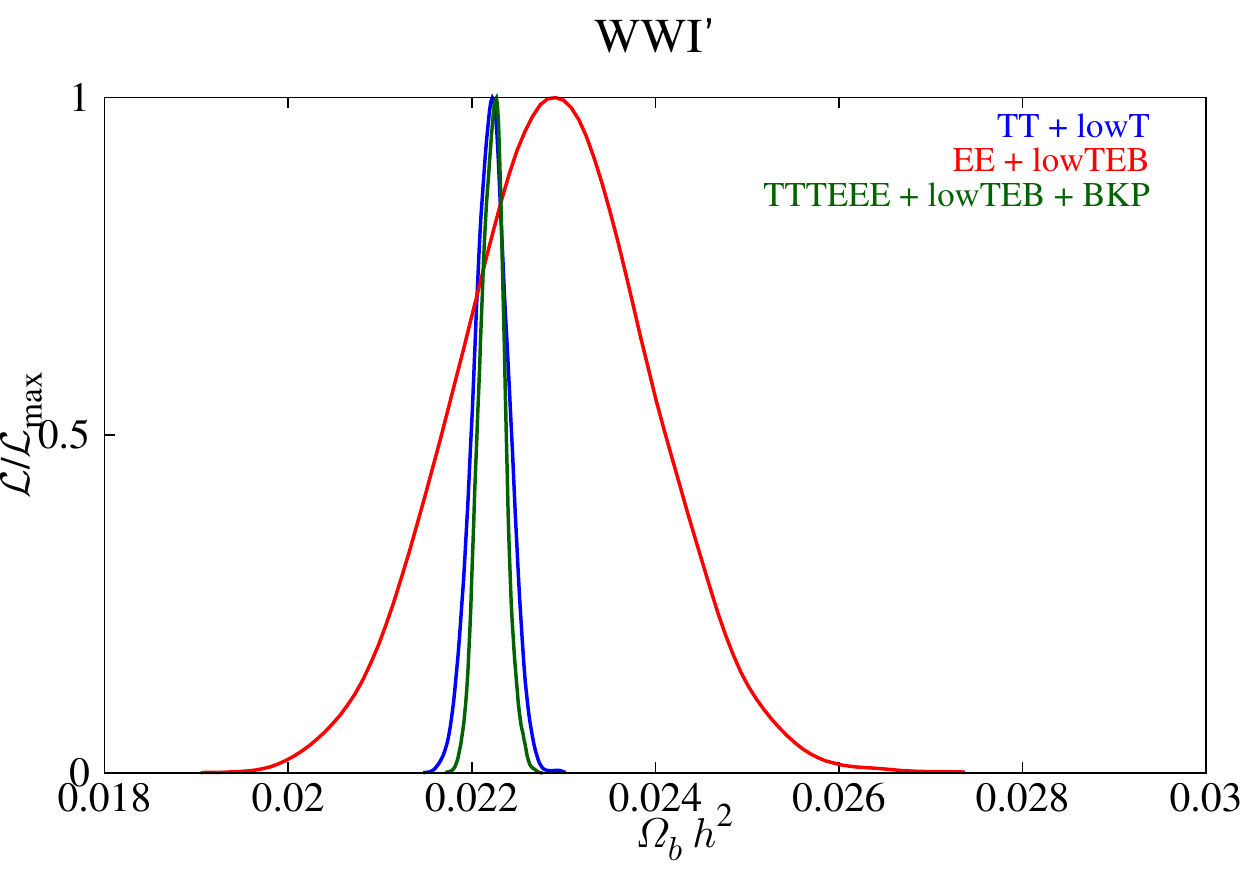}} 

\resizebox{142pt}{105pt}{\includegraphics{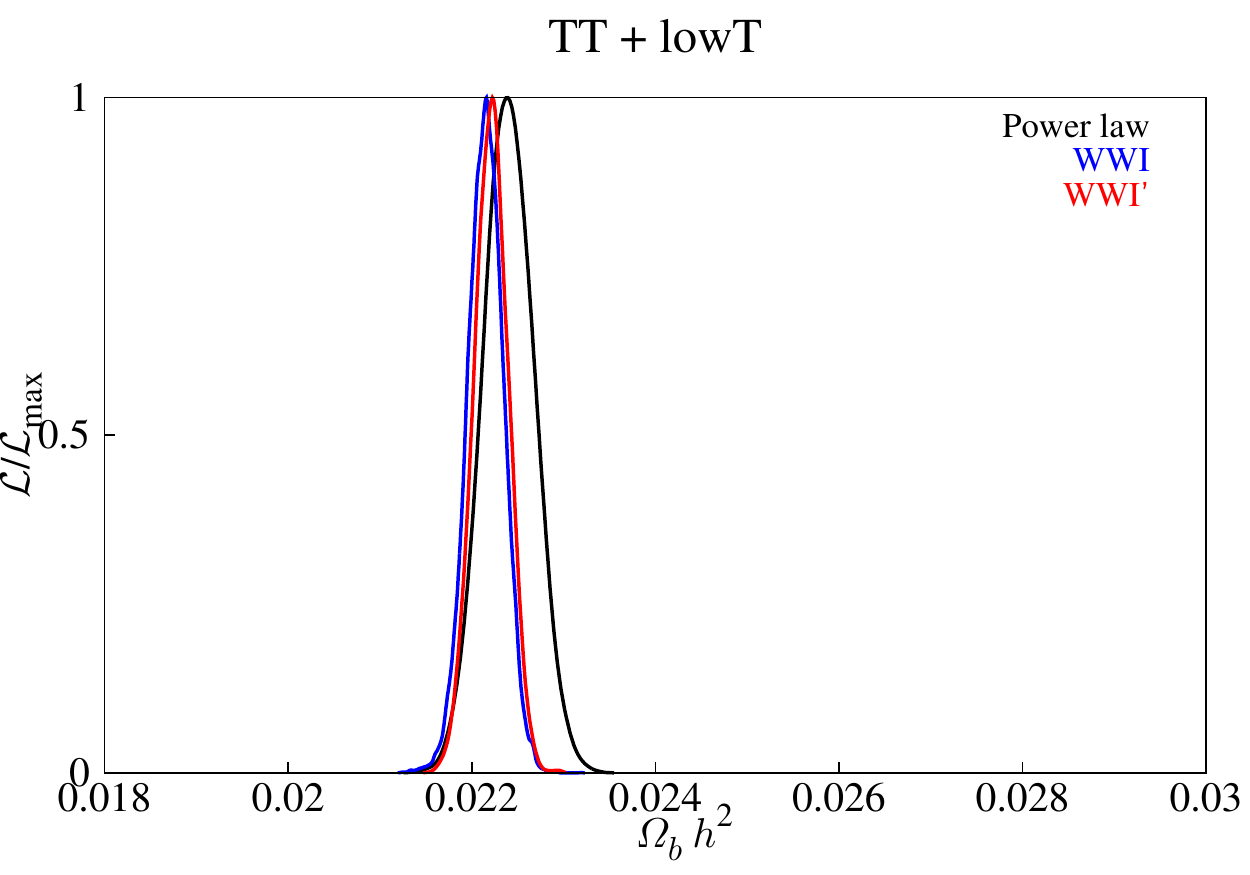}} 
\resizebox{142pt}{105pt}{\includegraphics{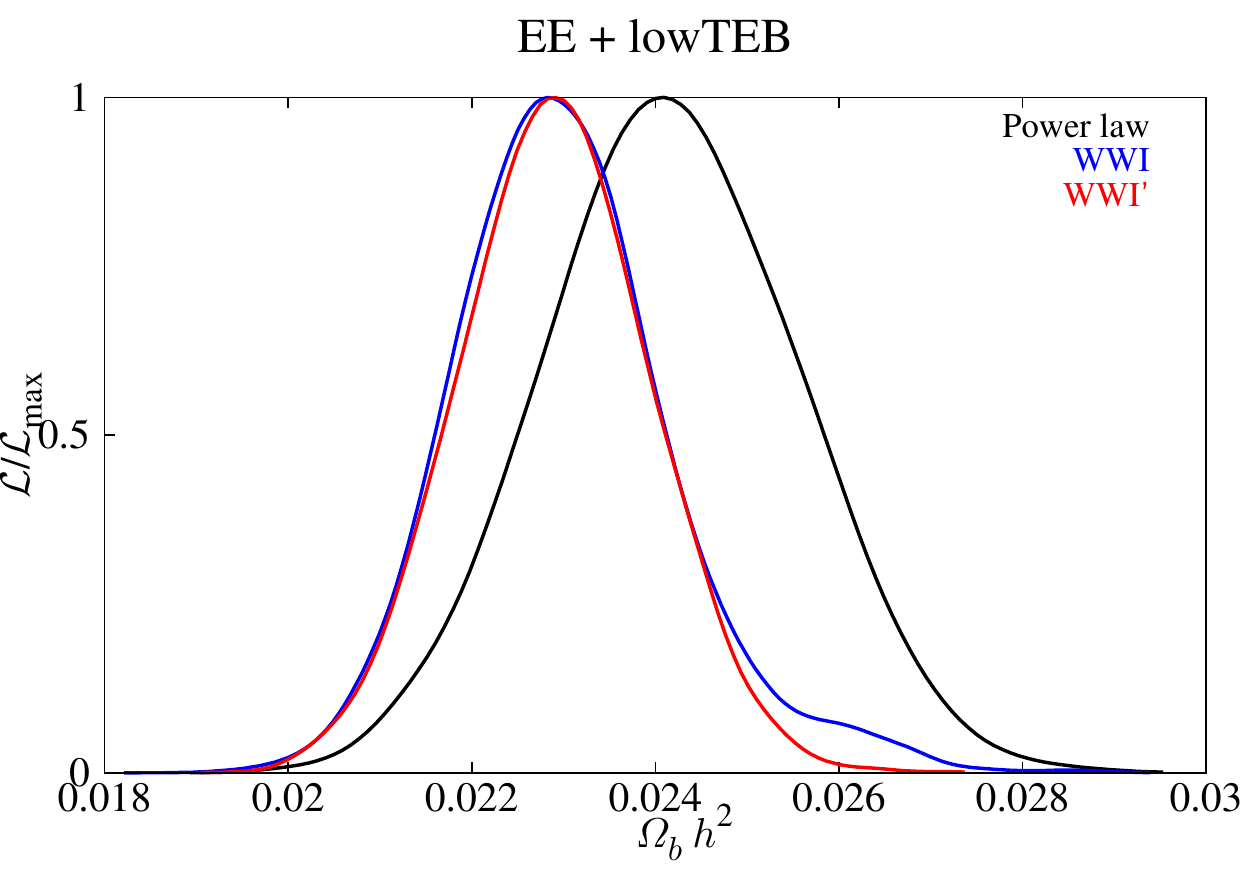}} 
\resizebox{142pt}{105pt}{\includegraphics{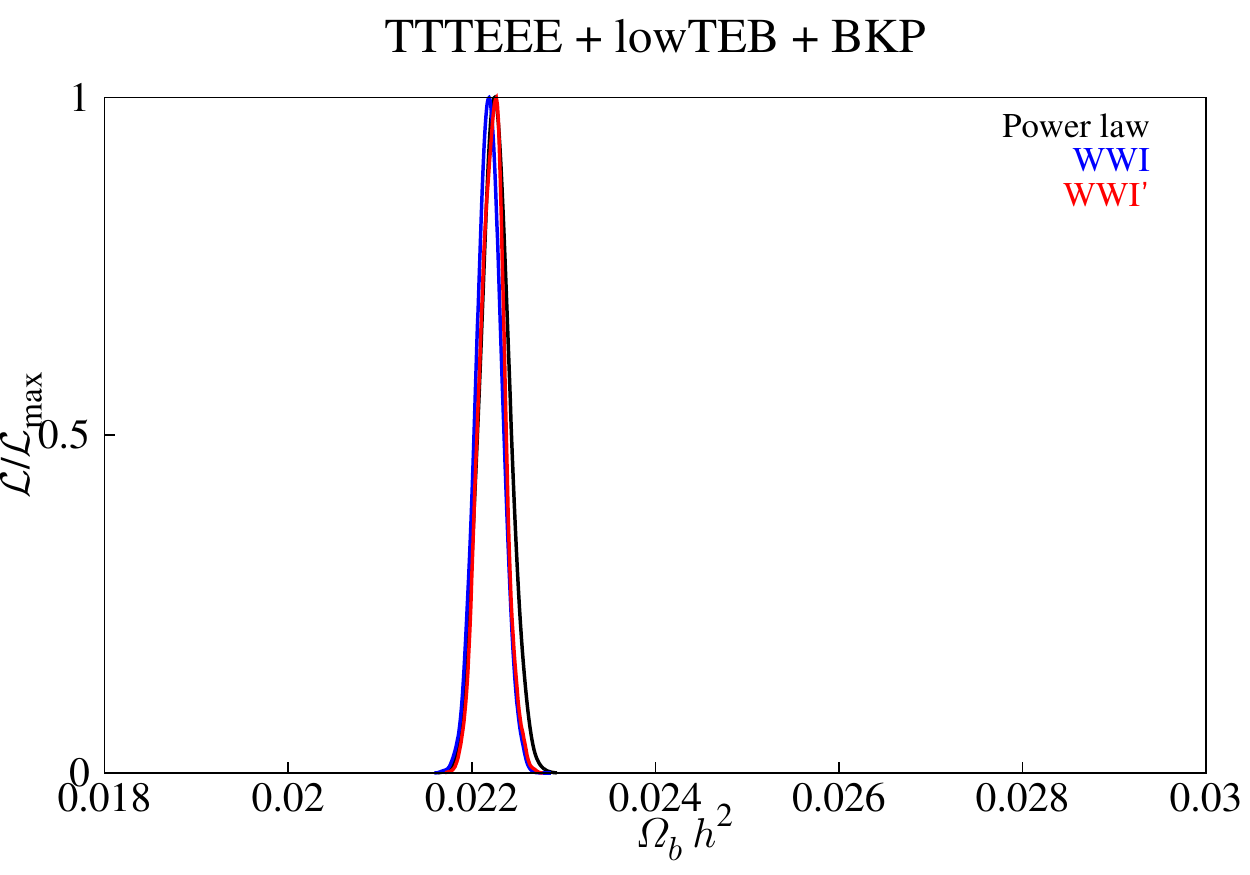}} 
\end{center}
\caption{\footnotesize\label{fig:omegabh2}The likelihood of baryon density obtained from temperature and polarization data from Planck-2015. In the top panel, in each plot,
we provide the likelihood obtained assuming the power law, WWI and WWI$'$ form of inflation. Note that when we assume power law PPS, the E polarization data favors a large baryon 
density ($\Omega_{\rm b}{h^2}\sim 0.024$) which looks having slight disagreement with the temperature data. WWI and WWI$'$ reduce this disagreement with a mean value of baryon density 
$\Omega_{\rm b}{h^2}\sim 0.023$. In the bottom panel we compare the results from different inflationary models and power law PPS for different datasets. The shift of the likelihoods
to the lower baryon density is evident for WWI and WWI$'$ models for EE + lowTEB data.}
\end{figure*}

The presence of features change the background parameter constraints from power law. Due to the features in WWI and WWI$'$ we find certain changes in the background
parameters. The power suppression at large angular scales shifts the reionization optical depth to a higher value $\tau\sim0.07-0.1$ compared to the power law case. 
Here, we present only the change in the baryon density constraints. In Fig.~\ref{fig:omegabh2}, we plot the marginalized likelihoods from power law, WWI and WWI$'$. 
Note that for WWI, in arriving at the constraints, we have fixed the width of the transition ($\Delta$) to the best fit value of WWI-d using the prior knowledge
that WWI-d is able to provide an agreement to the baryon density mismatch from power law PPS. In this figure the each plot at the top corresponds to different models
that compares the difference in the likelihood of baryon density from T, E and from their combination. For power law model we note that the peak of the likelihood
from EE + lowTEB is significantly far away from the tail of TT + lowT data. When feature models are used for the PPS, we find that the likelihoods shift towards lower 
baryon density for EE (from $\Omega_{\rm b}{h^2}\sim 0.024$ for power law to $\sim0.023$ for WWI and WWI$'$). In the bottom panel of the same figure, we provide 
the same likelihoods but plotted for T, E and complete datasets for comparison. We find that the shift is not substantial given the large standard deviation of EE 
data but it definitely reduces the tension to a fair extent that amounts to more than 13 improvement in $\chi^2$ values. Furthermore, we find that the tension is not relaxed 
due to the increase in degeneracy owing to the extra parameters. In fact, we find that for WWI and WWI$'$ models, the likelihood of baryon density is sharper than the 
power law. The standard deviation of the likelihood in WWI and WWI$'$ models are 0.0012 and 0.001 respectively while for power law it is 0.0014. Qualitatively the 
likelihoods can also be compared in middle plot of lower panel of Fig.~\ref{fig:omegabh2}. Here, we would like to point out that since our slow roll part of the potential generates a spectral tilt
$\sim 0.96$ (the asymptotic or average $n_{\rm S}$ at small scales), in our analyses we do not have the scope of marginalizing over the spectral tilt and hence the comparison
with power law case, where we marginalize over $n_{\rm S}$, might not be complete. Hence we obtained the standard deviation of the baryon density from the power law model 
against EE + lowTEB dataset by fixing the $n_{\rm S}=0.964$, which turns out to be  $\sim0.001$, similar to WWI$'$ model. For the complete datasets used in the analyses, we find that the standard
deviations for WWI and WWI$'$ are similar to the power law case as well. In other words, we find tighter (or at least similar) constraints on background parameters with WWI and 
WWI$'$ models as in power law model.

Though we are getting substantial improvement in likelihood compared to power law with both WWI and WWI$'$, we fail to get more than 1$\sigma$ evidence for these features. The maximum 
likelihoods show significant deviation from featureless case in both the models, but when marginalized, we find that the best fits have low marginalized probabilities. Hence, with Planck 
data we are unable to rule out featureless primordial power spectrum even though feature models agree with the data substantially better. 

At this point we should mention that WWI-[b,c,d] and WWI$'$ contain features around the BAO scales and hence the matter power spectra can be constrained using the 
large scale structure data as well. With the upcoming data from DESI~\cite{DESI}, SDSS-IV, e-BOSS~\cite{SDSSIV,eboss}, Euclid~\cite{Euclid} we expect to 
verify the existence of these features. Also with the upcoming low-$\ell$ EE data from Planck HFI is supposed to provide stronger constraints on the large scale suppression. We expect to
revisit Wiggly Whipped Inflation with the future datasets from CMB and LSS.

\subsection{Non-Gaussianity}
\begin{figure*}[!htb]
\begin{center} 
\resizebox{180pt}{100pt}{\includegraphics{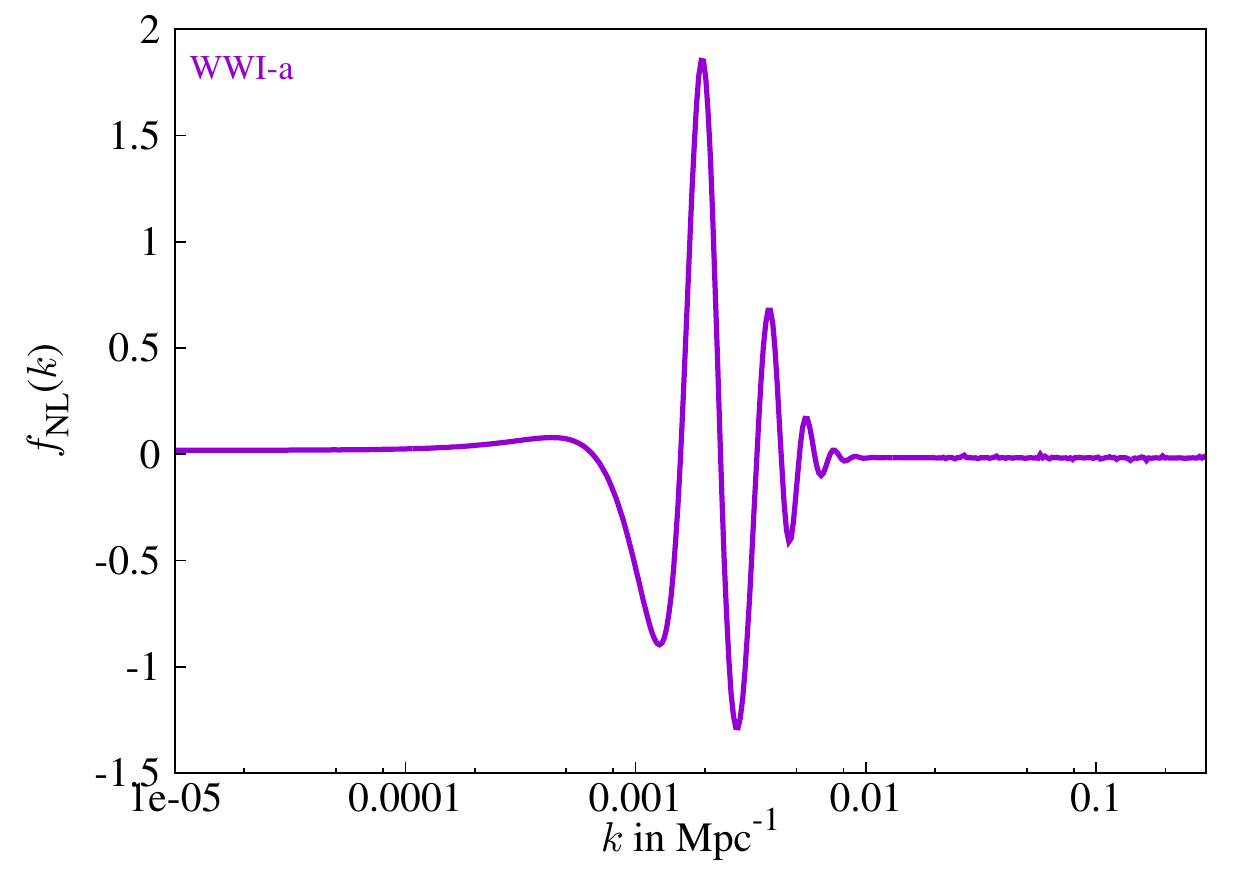}} 
\resizebox{180pt}{100pt}{\includegraphics{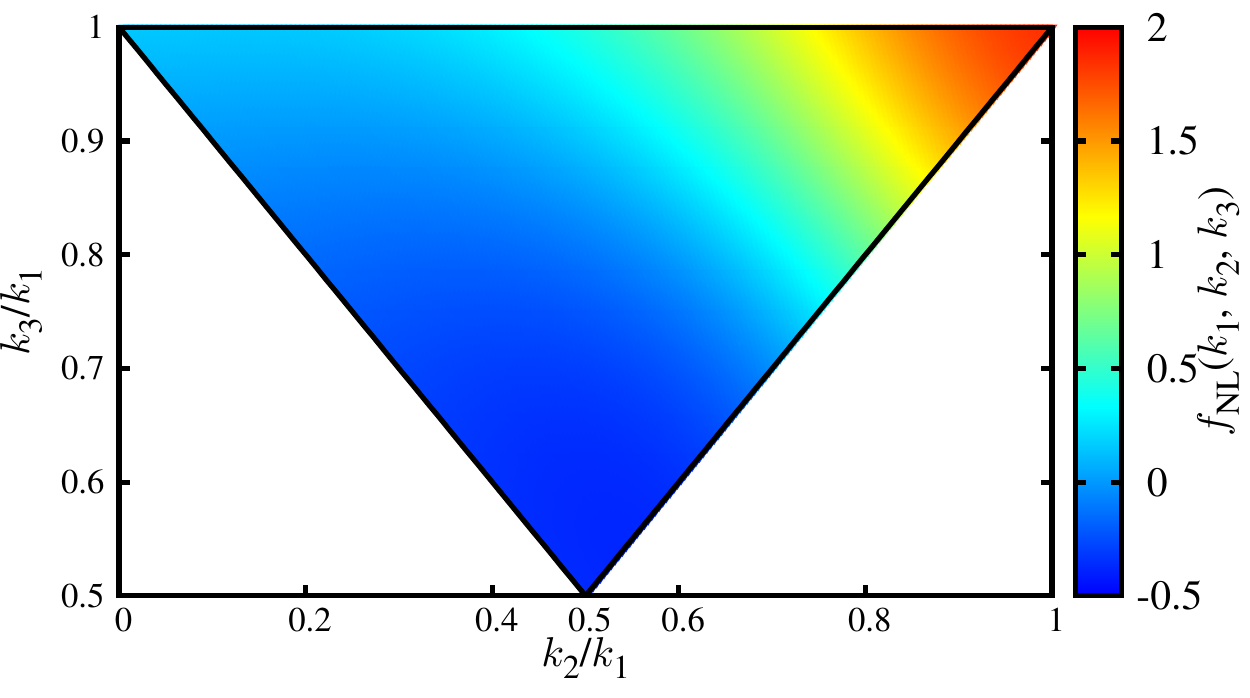}} 

\resizebox{180pt}{100pt}{\includegraphics{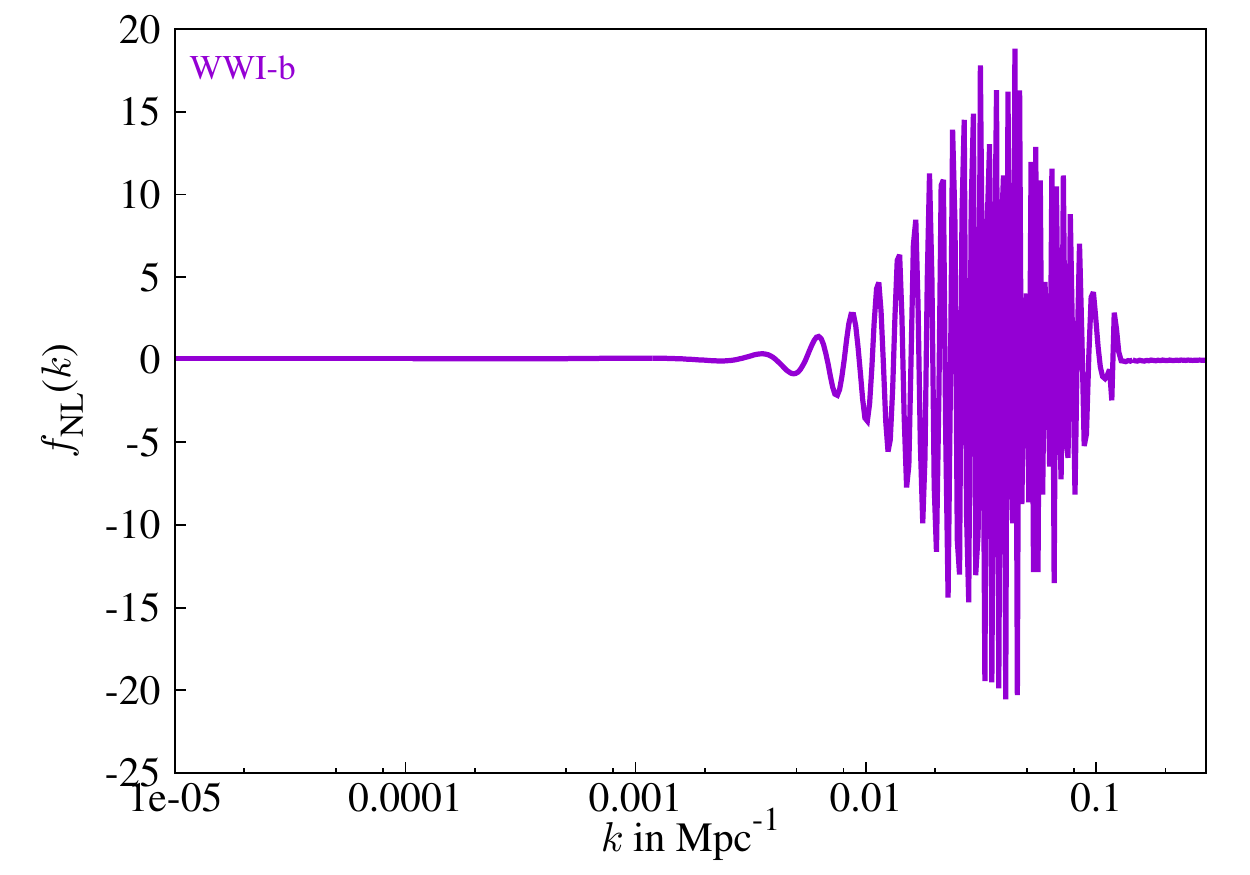}} 
\resizebox{180pt}{100pt}{\includegraphics{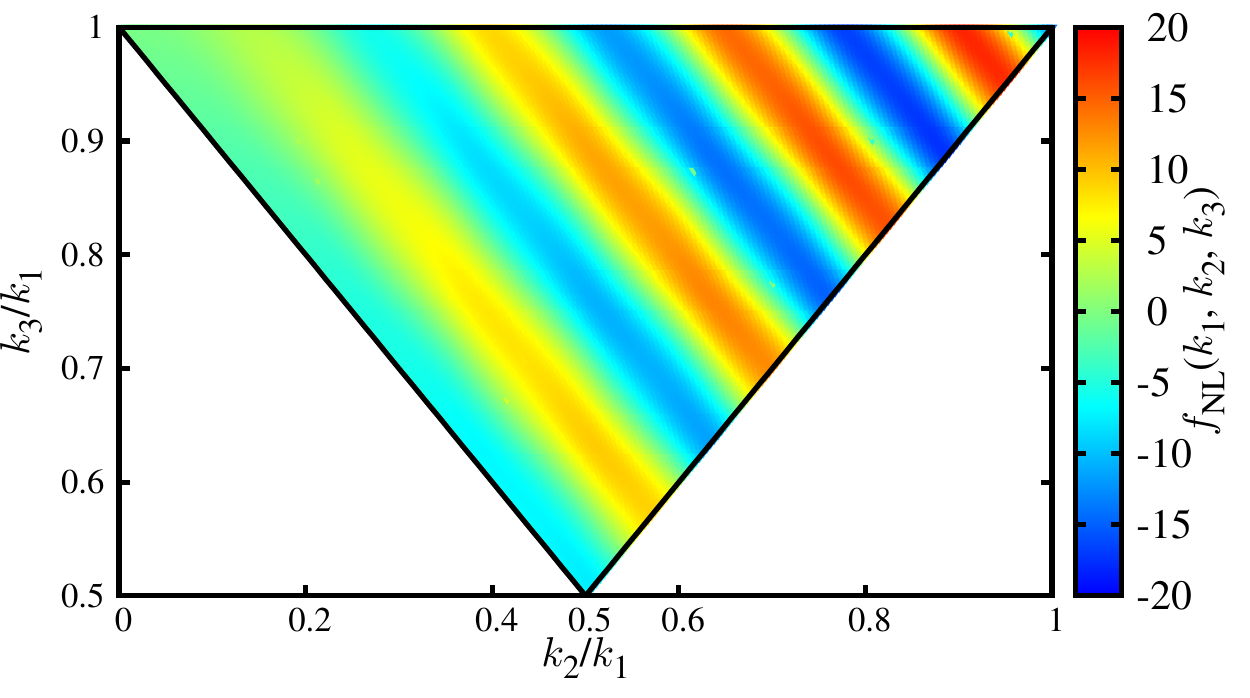}} 

\resizebox{180pt}{100pt}{\includegraphics{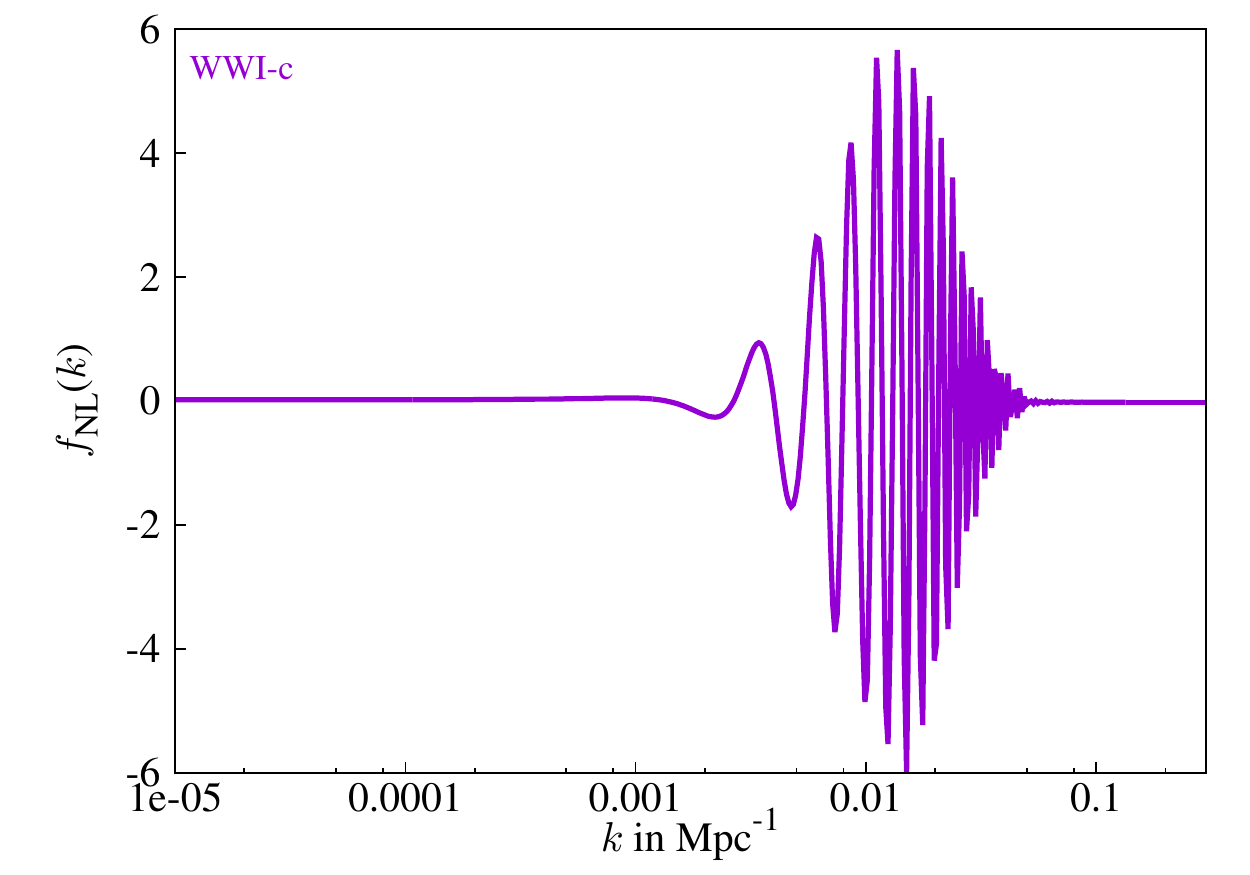}} 
\resizebox{180pt}{100pt}{\includegraphics{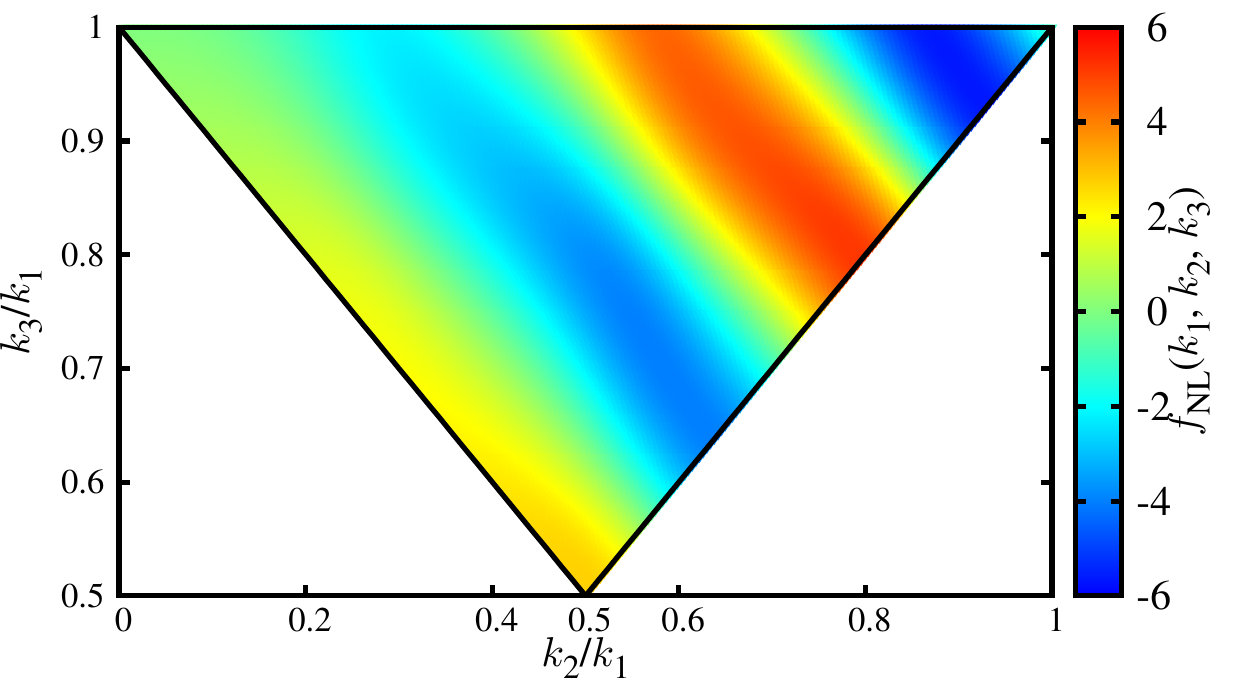}} 

\resizebox{180pt}{100pt}{\includegraphics{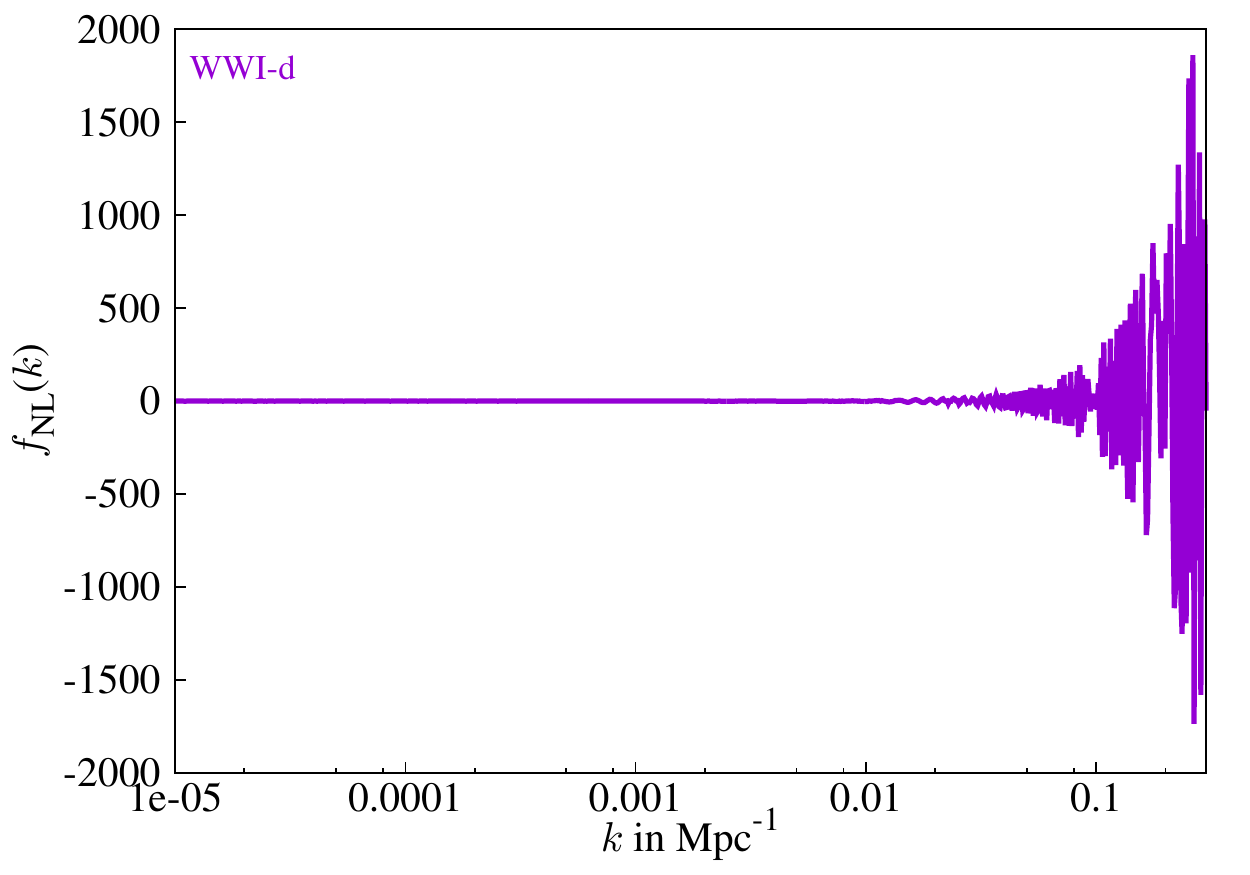}} 
\resizebox{180pt}{100pt}{\includegraphics{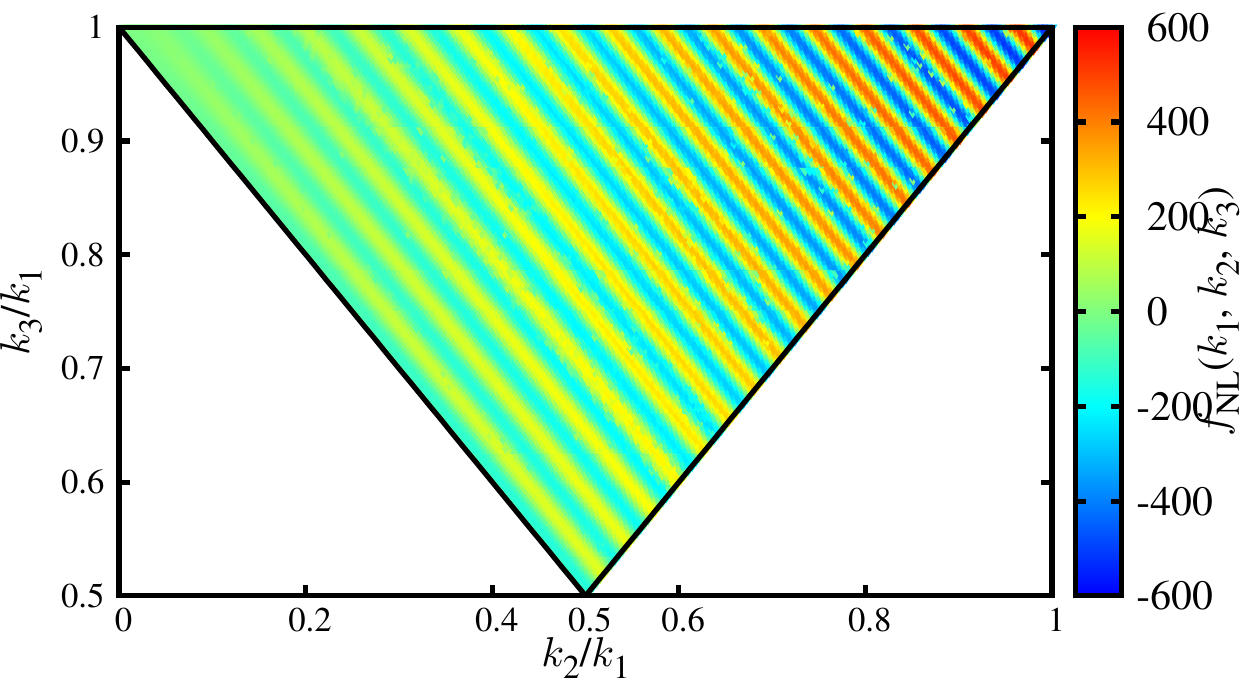}} 

\resizebox{180pt}{100pt}{\includegraphics{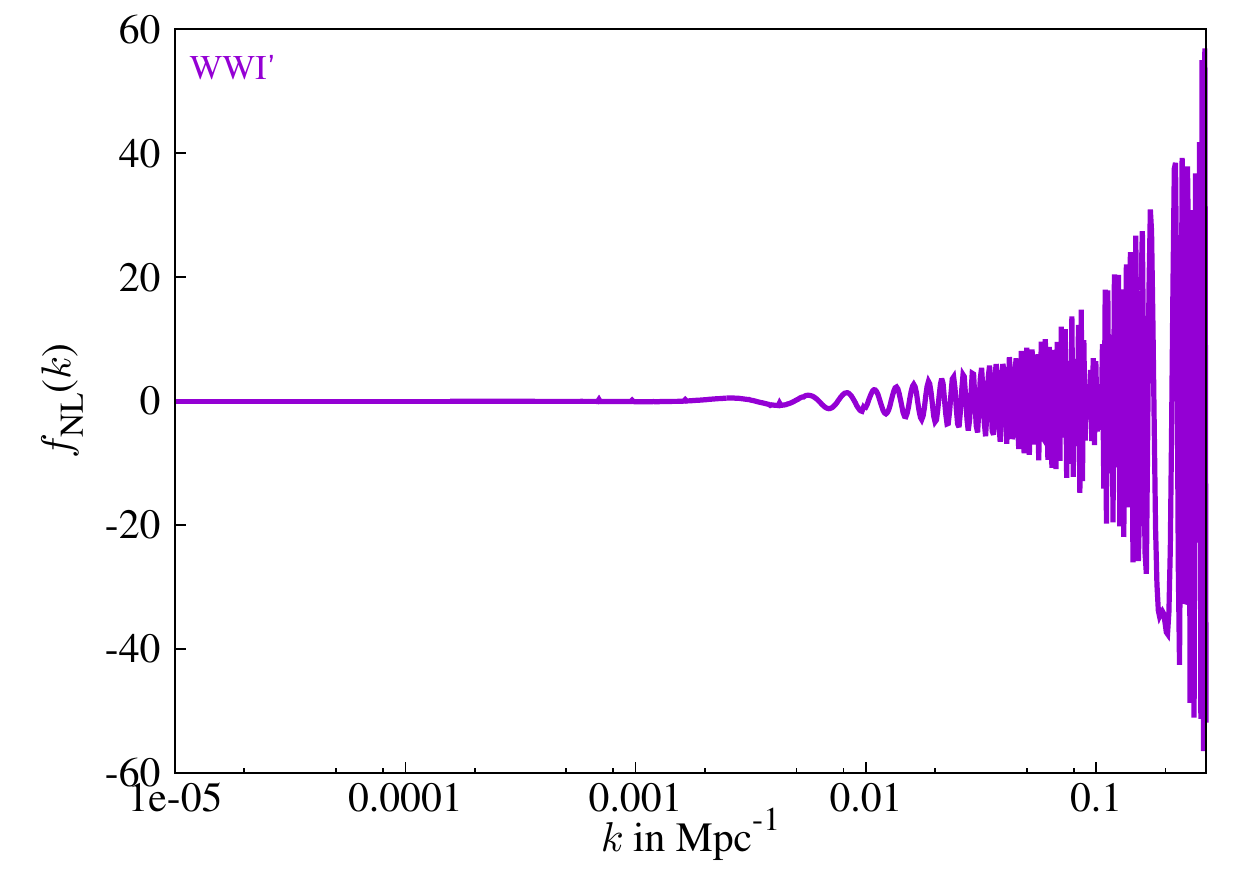}} 
\resizebox{180pt}{100pt}{\includegraphics{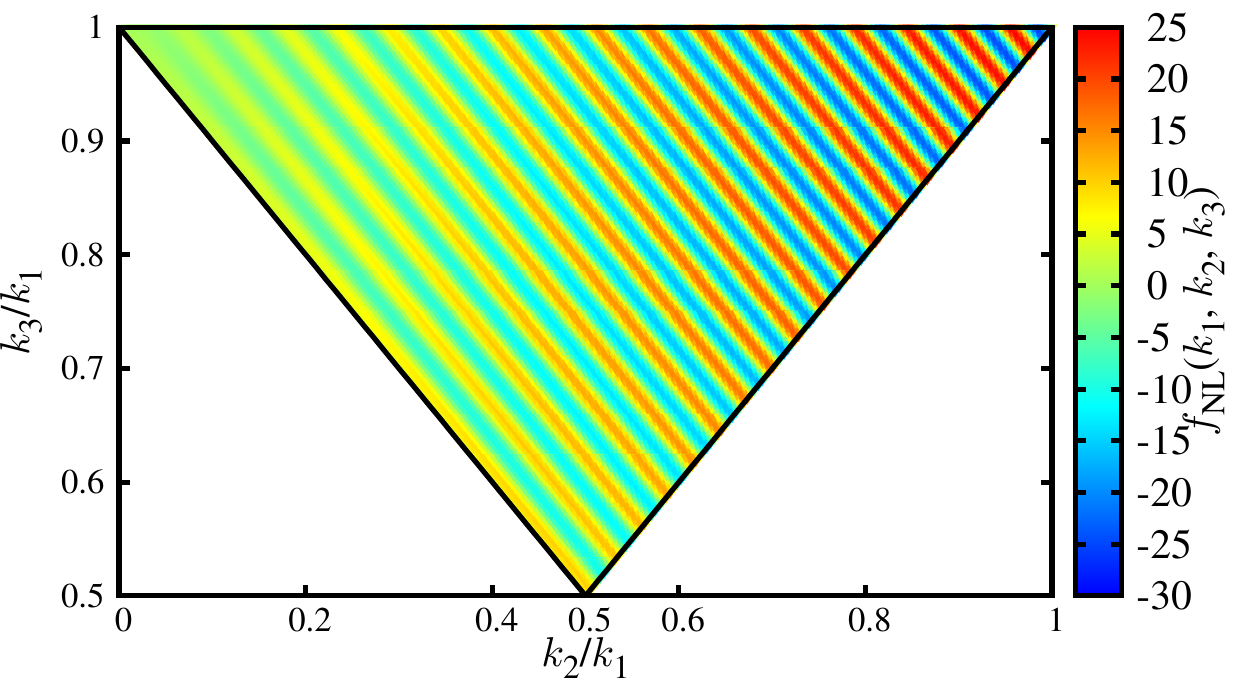}} 
\end{center}
\caption{\footnotesize\label{fig:fnl} Wiggly Whipped Inflation : $f_{\rm NL}$ in equilateral (left) and in arbitrary triangular configurations (right). From top to
bottom we plot the $f_{\rm NL}$ from WWI-a, WWI-b, WWI-c, WWI-d and WWI$'$ respectively (see, Fig.~\ref{fig:psk} for corresponding PPS). Note that while WWI-a provides $f_{\rm NL}\sim{\cal O} (1)$ which has a localized 
feature around $k\sim0.002~{\rm Mpc^{-1}}$, features that extend over wide range of cosmological scales with larger frequency, 
generate higher non-Gaussianity with $f_{\rm NL}\sim{\cal O} (100-1000)$.}
\end{figure*}

It has been extensively discussed in the literature that inflation models, that offers departure from slow roll inflation to generate features in the PPS,
also produce non-negligible non-Gaussianities~\cite{chen,Flauger-amm,Martin:2011sn,Hazra:2012BINGO,Sreenath:2014BINGO2,Arroja:2011,Arroja:2012,Martin:2014}. 
In this paper we shall only consider the three point 
correlations, {\it i.e.} the bispectra generated by WWI and WWI$'$. We use {\tt BINGO-2.0}~\cite{Hazra:2012BINGO,Sreenath:2014BINGO2,BINGOPAGE} to calculate the bispectra 
for our best fit models. We have used the best fit potential parameters
from TTTEEE + lowTEB + BKP for WWI-[a,b,c,d] and WWI$'$ and compute the $\fnl$. We calculate all the terms contributing to the $\fnl$ arising from the interaction Hamiltonian,
cubic in order of the curvature perturbations~\cite{maldacena-2003,chen,Martin:2011sn}. Since WWI best fit PPS do have features with high frequency 
and high amplitude, any slow-roll approximation in the 
bispectrum integral shall underestimate the actual value of the $\fnl$. Using {\tt BINGO}, we do not make any approximation and evaluate the bispectrum integral 
numerically. We calculate local $\fnl$ for our models derived from bispectrum $\cB_{_{\mathrm{S}}}(\vka,\vkb,\vkc)$,
\begin{eqnarray}
\fnl(\vka,\vkb,\vkc)=-\frac{10}{3}\, (2\,\pi)^{-4}\; (2\,\pi)^{9/2}\;
k_{1}^3\, k_{2}^3\,k_{3}^3\; \cB_{_{\mathrm{S}}}(\vka,\vkb,\vkc)\nn\\
\times\l[k_1^{3}\; {P}_{_{\mathrm{S}}}(k_2)\; 
{P}_{_{\mathrm{S}}}(k_3)
+{\mathrm{two~permutations}}\r]^{-1},
\end{eqnarray}
where ${P}_{_{\mathrm{S}}}(k)$ denotes the primordial power spectra.

In Fig.~\ref{fig:fnl} we plot the $\fnl$ for the best fit potentials obtained from TTTEEE + lowTEB + BKP datasets. To the left we plot the $\fnl$ in equilateral
limit ($k_1=k_2=k_3=k$). To the right we plot the 2D heat map of the $\fnl$. The 2D $\fnl$ are plotted as a function of $k_3/k_1$ and $k_2/k_1$. The top left corner of 
the triangular configurations represent the squeezed limit ($k_2,k_3<<k_1$) and the top right corner represent the equilateral limit. $k_1$ in WWI-[a,b,c,d] and WWI$'$ are chosen 
to be $\sim2\times10^{-3}$, $0.03$, $0.015$, $0.14$ and $0.14$ ${\rm Mpc^{-1}}$ respectively. For the first three cases, we chose the mode $k_1$ by locating the scale where $\fnl$ becomes
maximum and for WWI-d and WWI$'$ we chose it to be a smaller scale since for a sharp transition in the potential (or in its derivative in the latter case), the $\fnl$ is expected to diverge 
linearly with wavenumber and a smaller $k_1$ is expected to capture the profile of the $\fnl$ in the 2D heat map. 

Note that the WWI-a and WWI-c generates $|\fnl|\sim1-10$. Because of the presence of a broad step in the potential and wide frequency oscillations, the slow-roll parameters 
$\epsilon_{i+1}=\d \ln \epsilon_{i+1}/\d N$ and their derivatives are not large enough
to produce large three point correlations of curvature perturbations. While WWI-b, WWI-d and WWI$'$ generates higher $\fnl$ because of sharper transition from moderate fast roll to 
slow-roll potential ($\Delta \rightarrow 0$). We would like to point a crucial difference between WWI-b and WWI-d (or WWI$'$) models. When we compare these two models with the temperature and polarization
angular power spectrum, we find similar improvement in fit compared to power law models. Hence, to the power spectra level, these models are nearly indistinguishable. On the other hand
the WWI-b generates $\fnl$ which converges at small scales but WWI-d and WWI$'$ diverges with the wavenumbers because the singularity at $\phi=\phi_{\rm T}$ (although, note that in strict 
numerical sense, we have to model the singularity by a transition width very close to zero). The divergent $\fnl$ arising from the instantaneous transition and the effect of smoothing 
the discontinuity were studied before in literature~\cite{Arroja:2011,Arroja:2012,Martin:2014}\footnote{We find $\Delta\sim10^{-3}$ for WWI-b and $10^{-5}$ for WWI-d}.
As we have pointed out before, the WWI$'$ essentially generates same power spectra as generated by Starobinsky-1992 model of inflation with a spectral tilt of $\sim0.96$.
The bispectra generated from this model, hence are of same shape as discussed in literature~\cite{Arroja:2011,Martin:2011sn,Arroja:2012,Martin:2014,Sreenath:2014BINGO2}.
To evaluate the $\fnl$ from the WWI$'$ model we have used a Theta function and Delta function with a width similar to WWI-d model, to reproduce the effects of the singularity in the 
second derivative of the potential. We should emphasize that an instantaneous transition is not a realistic situation and there has to be a finite width associated with the 
transition in the potential that ensures the convergence of bispectra at small scales (as have been emphasized in~\cite{Martin:2014}). 

From the analyses above we can state that feature models, that are indistinguishable from the likelihood {\it w.r.t.} the power spectra data, a joint estimation of PPS and 
bispectra can quantitatively be able to distinguish between the models, and also provide extra significance for primordial features in the data, if present.
There are hints of oscillatory bispectra from Planck 2015 analysis~\cite{Planck:2015NG} and hence it is important to confront the WWI features with Planck bispectra data.
We emphasize that since WWI, within its single framework, offers a wide variety of features, this model will be extremely useful in comparing different features.
In this context, note that, the models considered in our paper, oscillations in the PPS at large $k$ do not play a significant role for the explanation of the features 
in the CMB multipole spectrum apart from WWI-d. This is in agreement with the conclusion of~\cite{Fergusson:2014tza} that there is no statistically significant signs of 
sharp oscillatory features in the CMB spectrum and bispectrum.

\section{Conclusions}\label{sec:conclusions}
We show that the WWI framework that was introduced mainly to explain the BICEP2 and the Planck 2013 results in a single theoretical model, can 
explain a wide variety of the primordial features that are obtained from direct reconstructions using CMB angular power spectrum data and also that 
are motivated from high energy theories. Without using different theoretical models of inflation, WWI allows us to generate different types of PPS features
in a single framework, making it extremely suitable for primordial feature hunt and also in constraining background cosmological parameters marginalized over 
a multitude of inflationary scenarios.
In this paper we confront Wiggly Whipped inflaton potentials allowing deviations from strict slow roll, against the latest Planck 2015 angular power 
spectrum data of temperature and polarization and BICEP2/Keck B-mode polarization data.
WWI offers a simple transition from moderate fast roll to strict slow roll potential with/without the presence of a discontinuity/jump in the potential at the field value of phase transition.
We also present WWI$'$, in which a smooth part of the inflaton potential is that of the $R+R^2$ inflationary model~\cite{Starobinsky:1980te} in the Einstein frame (and which in turn represents a 
special case of the $\alpha$-attractor model~\cite{alpha_attractor}) and discontinuity appears only in the first derivative of the potential. 
We have been able to generate a wide class of primordial features that have been discussed in the literature within the frameworks of WWI and WWI$'$. Owing to the flexibility of the 
WWI model, we are able to locate the local minima and possibly the global minima in the parameter space of WWI using only one potential. From the individual and joint analyses of T, E and B 
polarization we identify 4 distinguishable features in the primordial power spectra of WWI, namely WWI-[a,b,c,d]. In all the 
cases we notice a common pattern, the large scale suppression of scalar perturbation spectra. WWI-a provides a dip in the power spectra at $2\times10^{-3}{\rm Mpc^{-1}}$, which has been 
discussed extensively in the literature and we find that this feature, apart from providing an improved fit to the low-$\ell$ likelihood (mostly from around $\ell=22$), also provides 
better fit to high-$\ell$ EE datasets. We find localized oscillations (but wider than WWI-a) within $0.002-0.05~{\rm Mpc^{-1}}$ that particularly agree with high-$\ell$ EE data
compared to power law best fit. WWI-a and WWI-c represent the primordial features that attempts to address the features in the individual CMB angular power spectra data, that are not addressed 
by standard power law PPS. Apart from these best fits, our analysis with WWI offers a third kind of feature where the primordial power spectrum do not provide notable improvement 
in fit when compared with TT and EE angular power spectrum individually, but provides more than 13 improvement in fit in $\chi^2$ compared to power law {\it w.r.t.} the complete dataset. Similarly 
in WWI$'$, using only 2 extra parameters in the inflation potential, we find $\sim12$ improvement in $\chi^2$ fit. WWI$'$ offers a step like suppression at larger scales accompanied by oscillations
at smaller scales. In this model we show that primordial feature of a fixed shape, by changing its location and amplitude, can match both temperature, polarization data separately and also in a combined analysis. 
We find that the standard baseline model of cosmology prefers higher baryon density ($\Omega_{\rm b}h^2\sim0.024$ as best fit value) for E-mode polarization which is more than 3$\sigma$ away from the TT best fit. 
Though the uncertainties in the E-mode polarization data are much larger than temperature data and the distance to the best fit values from EE data can not be trusted in a statistically robust analysis, 
we demonstrated that TT, EE Likelihood decrease in a joint analysis.~\footnote{Here we would like to mention that this difference can also be an artifact of the systematics in the Planck polarization and 
temperature data~\cite{Addison:2015wyg}} WWI-b, WWI-d and WWI$'$ offering high frequency oscillations extending over a large range of cosmological scales 
offer a scenario that fits the TT, EE, EE and lowTEB data better in joint analyses, keeping the best fit value of $\Omega_{\rm b}h^2\sim0.0222$, as demanded by temperature 
power spectrum. Both WWI and WWI$'$ show a shift in baryon density to a lower value ($\Omega_{\rm b}h^2\sim0.023$) when compared with EE dataset. 
We find that WWI-b and WWI-d represent the global best fit to the complete Planck 2015 and BICEP2/KECK datasets. The fundamental difference between WWI-b and WWI-d is in the 
sharpness of their transition from moderate fast roll to the complete slow roll regime. These two features are 
indistinguishable from the angular power spectrum analyses but we show that they have very distinct bispectra signatures. It is possible to find stringent constraints on the WWI and WWI$'$ models 
upon joint analyses with CMB power spectra and bispectra data.


\section*{Acknowledgments}
DKH and GFS acknowledge Laboratoire APC-PCCP, Universit\'e Paris Diderot and Sorbonne Paris Cit\'e (DXCACHEXGS)
and also the financial support of the UnivEarthS Labex program at Sorbonne Paris Cit\'e (ANR-10-LABX-0023 and ANR-11-IDEX-0005-02).
DKH would like to thank the hospitality of Cluster Computing Center (through the support from the DOE HEP's Forum on Computational Excellence) 
and Berkeley Center for Cosmological Physics, LBL, Berkeley and Princeton University where a part of the work has been carried out. AS would like to acknowledge
the support of the National Research Foundation of Korea (NRF-2016R1C1B2016478). AAS was partially supported by the grant RFBR 14-02-00894 and by the Russian 
Government Program of Competitive Growth of Kazan Federal University.

\end{document}